\documentclass[12pt]{article}
\pdfoutput=1
\usepackage{amssymb,amsmath}
\numberwithin{equation}{section}
\usepackage{epsfig}
\usepackage{epstopdf}
\usepackage{latexsym}
\usepackage{graphicx}
\usepackage{booktabs}
\usepackage{bbm}
\usepackage{enumitem}
\usepackage[numbers,compress]{natbib}
\usepackage{bm}
\usepackage[T1]{fontenc}
\usepackage{amsthm}

\usepackage[margin=20pt,small]{caption}
\usepackage{subcaption}

\usepackage[toc]{appendix}

\usepackage{color}
\usepackage{datetime}
\usepackage[
      colorlinks=false,
      linkcolor=darkblue,  
      urlcolor=blue,    
      filecolor=blue,     
      citecolor=red,
linktocpage=true,
      pdfstartview=FitV,
      bookmarksopen=true,
	  hidelinks
      ]{hyperref}

\ifpdf
\fi

\newcommand*{\boxedcolor}{red}
\makeatletter
\renewcommand{\boxed}[1]{\textcolor{\boxedcolor}{%
  \fbox{\normalcolor\m@th$\displaystyle#1$}}}
\makeatother

\usepackage{calligra}
\DeclareMathAlphabet{\mathcalligra}{T1}{calligra}{m}{n}

\definecolor{cardinal}{rgb}{0.6,0,0}
\definecolor{darkgreen}{rgb}{0,0.5,0}
\definecolor{golden}{rgb}{0.92, 0.7, 0}
\definecolor{midnight}{rgb}{0, 0, 0.5}
\definecolor{darkblue}{rgb}{0.2, 0, 0.8}

\def\cB{{\cal B}}

\def\cI{{\cal I}}

\def\cN{{\cal N}}

\def\cV{{\mathcal V}}

\def\Tr{{\rm Tr}\,}

\def\fg{\frak{g}}

\def\fn{\frak{n}}

\newcommand{\be}{\begin{equation}} \newcommand{\ee}{\end{equation}}
\newcommand{\bea}{\begin{equation} \begin{aligned}} \newcommand{\eea}{\end{aligned} \end{equation}}
\newcommand{\bmu}{\begin{multline}} \newcommand{\emu}{\end{multline}}

\newcommand{\ie}{\textit{i.e.}}

\newcommand\equ[1] {\begin{equation}#1\end{equation}}

\newcommand\eqs[1] {\begin{align}#1\end{align}}
\newcommand\eqss[1] {\begin{align}\begin{split}#1\end{split}\end{align}} 
\newcommand\eqsn[1] {\begin{align*}#1\end{align*}}
\renewcommand\( {\left(}
\renewcommand\) {\right)}

\newcommand{\dvol}{d\mathrm{vol}}
\newcommand{\vol}{\mathrm{vol}}
\usepackage[cmtip,all]{xy}
\newcommand{\longsquiggly}{\xymatrix{{}\ar@{~>}[r]&{}}}
\begin{document}  
%%%%%%%%%%%%%%%%%%%%%%%%%%%%%%

\begin{titlepage}
 
\medskip
\begin{center} 
{\Large \bf  A Universal Counting  of  Black Hole Microstates in AdS$_4$}

\bigskip
\bigskip
\bigskip
\bigskip
\bigskip

{\bf Francesco Azzurli,${}^{(1)}$ Nikolay Bobev,${}^{(2)}$ P. Marcos Crichigno,${}^{(3)}$ Vincent S. Min,${}^{(2)}$ and  Alberto Zaffaroni${}^{(1)}$  \\ }
\bigskip
\bigskip
${}^{(1)}$ Dipartimento di Fisica, Universit\`a di Milano-Bicocca, I-20126 Milano, Italy\\
INFN, sezione di Milano-Bicocca, I-20126 Milano, Italy\\
\vskip 5mm
${}^{(2)}$
Instituut voor Theoretische Fysica, KU Leuven \\
Celestijnenlaan 200D, B-3001 Leuven, Belgium
\vskip 5mm
${}^{(3)}$ Institute for Theoretical Physics, University of Amsterdam,\\ 
Science Park 904, Postbus 94485, 1090 GL, Amsterdam, The Netherlands
\vskip 5mm
\texttt{francesco.azzurli@mib.infn.it,~nikolay.bobev@kuleuven.be,~p.m.crichigno@uva.nl,\\~vincent.min@kuleuven.be,~alberto.zaffaroni@mib.infn.it} \\
\end{center}

\bigskip
\bigskip

\begin{abstract}

\noindent  
\end{abstract}

\noindent 

Many three-dimensional $\mathcal{N}=2$ SCFTs admit a universal partial topological twist when placed on hyperbolic Riemann surfaces. We exploit this fact to derive a universal formula which relates the planar limit of the topologically twisted index of these SCFTs and their three-sphere partition function. We then utilize this to account for the entropy of a large class of supersymmetric asymptotically AdS$_4$ magnetically charged black holes in M-theory and massive type IIA string theory. In this context we also discuss novel AdS$_2$ solutions of eleven-dimensional supergravity which describe the near horizon region of large new families of supersymmetric black holes arising from M2-branes wrapping Riemann surfaces.

\end{titlepage}

%%%%%%%%%%%%%%%%%%%%%%%%%%%%%%%%%%%%%

\setcounter{tocdepth}{2}
\tableofcontents

\newpage

%%%%%%%%%%%%%%%%%%%%%%%%%%%%%%%%%%%%%
\section{Introduction}
%%%%%%%%%%%%%%%%%%%%%%%%%%%%%%%%%%%%%

String theory offers a unique vantage point into the physics of black holes. One the one hand it gives rise to supergravity theories in ten and eleven dimensions as its low energy limits and these theories admit a plethora of classical black hole (or more generally black brane) backgrounds. These black holes are sourced by the massive strings and branes, the microscopic degrees of freedom in the theory. On the other hand one can understand the low energy physics of these branes in terms of the quantum field theory living on their worldvolume. This dichotomy has been the source of many advances in our understanding of black holes, as well as being ultimately responsible for the concrete realization of the holographic principle in string theory. A famous example of this success story is the microscopic counting of the entropy for a class of asymptotically flat black holes in string theory \cite{Strominger:1996sh} (see also \cite{Sen:1995in} for earlier important work). However, this perspective on black holes has not been put to good use in the context of asymptotically AdS black holes. This state of affairs changed recently with the results in \cite{Benini:2015eyy}. The authors of  \cite{Benini:2015eyy} employed a certain supersymmetric index, $Z_{S^1\times \Sigma_{\mathfrak{g}}}$, defined in \cite{Benini:2015noa} to account for the entropy of a large class of supersymmetric asymptotically AdS$_4$ black holes arising from M2-branes wrapped on Riemann surfaces. These results were later extended to dyonic black holes in \cite{Benini:2016rke}. See also \cite{Hosseini:2016tor,Hosseini:2016ume,Benini:2016hjo,Closset:2016arn,Hosseini:2016cyf,Cabo-Bizet:2017jsl} for further recent developments.

Our goal in this work is to build upon this recent success in a number of ways. First, we show that there is a universal expression for the topologically twisted index in the large $N$ limit for a large class of $\mathcal{N}=2$ three-dimensional SCFTs.\footnote{The conditions on the $\mathcal{N}=2$ theory are discussed in detail in Section \ref{sec:fieldtheory}.} Namely, we find that the index is related to the three-sphere free energy, $F_{S^3}$, of the three-dimensional theory through the simple relation
\begin{equation}\label{eq:univintro}
\log Z_{S^1\times \Sigma_{\mathfrak{g}}} = (\mathfrak{g}-1) \, F_{S^3}\;.
\end{equation}
The universality of this relation stems from a universal partial topological twist that is used in the definition of the twisted index. In general the index depends on a set of background fluxes for the global symmetries of the $\mathcal{N}=2$ SCFT as well as a set of complex fugacities. A given choice of Riemann surface and background fluxes represents a particular topological twist of the three-dimensional theory on $\Sigma_\fg$. It was argued in \cite{Benini:2015eyy} that for a given such choice of twist one obtains the degeneracy of vacua  after extremizing the index as a function of the complex fugacities, and this extremal value appears in  \eqref{eq:univintro}. The universal topological twist, for which the relation in \eqref{eq:univintro} is valid, is singled out by choosing the background fluxes on $\Sigma_\fg$ such that the only non-zero flux is in the direction of the exact superconformal R-symmetry of the three-dimensional $\mathcal{N}=2$ SCFT.  

Such a simple universal relation in QFT should have an equally simple bulk realization for three-dimensional $\mathcal{N}=2$ SCFTs with a holographic dual. Indeed, we show that there is a simple black hole solution with a hyperbolic horizon in minimal four-dimensional gauged supergravity which provides the holographic realization of the relation \eqref{eq:univintro}. In fact, the entropy of this black hole, originally found in \cite{Romans:1991nq} and later studied further in \cite{Caldarelli:1998hg}, in the large $N$ limit is simply given by the twisted index and thus we arrive at a microscopic understanding of the entropy for this simple universal black hole. An important point in our story is the fact that this simple black hole solution can be embedded in eleven-dimensional and massive type IIA supergravity in infinitely many ways. This provides further evidence that the relation  \eqref{eq:univintro} holds (at least in the planar limit) for a large class of three-dimensional $\mathcal{N}=2$ SCFTs.

In the course of our analysis, we also clarify the relation proposed in \cite{Benini:2015eyy} between the topologically twisted index and the entropy of the black hole.  According to the holographic dictionary the logarithm of the partition function of the CFT  in the large $N$ limit should correspond to the on-shell action of the dual gravitational background. We show that, for the universal black hole, the on-shell action indeed coincides with (minus) the entropy. This computation is subtle and it should be done by considering a non-extremal deformation of the black hole and carefully
taking the extremal limit.

The string and M-theory embedding of the universal black hole can be interpreted as arising from D2- or M2-branes wrapped on $\Sigma_\fg$. As pointed out in \cite{Bershadsky:1995qy} these wrapped branes constructions implement the partial topological twist for the QFT which describes the low-energy dynamics of the branes. For a large number of branes one has to work in the supergravity limit in which one typically finds a black hole solution with a near horizon AdS$_2\times \Sigma_\fg$ geometry. This holographic construction was first employed for D3- and M5-branes in \cite{Maldacena:2000mw} and later generalized to M2-branes in \cite{Gauntlett:2001qs}.\footnote{For a review on branes wrapped on calibrated cycles and further references see \cite{Gauntlett:2003di}.} This prompts us to seek generalization of the simple black hole solutions in M-theory. Due to the complicated nature of the BPS supergravity equations in eleven-dimensional supergravity we focus only on finding explicit solutions for the AdS$_2$ near horizon region of the black hole. Employing this approach we find a large class of analytic explicit AdS$_2$ solutions of the eleven-dimensional supersymmetry equations with non-trivial internal fluxes. The full black hole solution for which these backgrounds are a near-horizon limit should correspond to turning on background magnetic fluxes for the global non-R symmetry in the dual CFT. The entropy of the black hole should then be captured by the twisted topological index with the given background global symmetry fluxes. While we do not establish fully this holographic picture we provide some evidence for its consistency. We should point out that the explicit AdS$_2$ solutions which we find fall into the general classification of such backgrounds in M-theory discussed in \cite{Kim:2006qu,Gauntlett:2007ts}. Other examples of AdS$_2$ vacua in eleven-dimensional supergravity of the class discussed here can be found in \cite{Gauntlett:2006ns,Kim:2012ek,Donos:2012sy,Kim:2013xza}.

We begin our story in the next section with a discussion of a universal partial topological twist that can be applied to three-dimensional $\mathcal{N}=2$ SCFTs and we argue how for a large class of these twisted theories one can calculate the topologically twisted index of \cite{Benini:2015noa} in the planar limit. We then proceed in Section \ref{sec:A simple 4d black hole} with a holographic description of the RG flow induced by the topological twist, which is realized by a black hole in four-dimensional minimal gauged supergravity. In addition, we show how to embed this black hole in various ways in M-theory and massive type IIA string theory. In Section \ref{AdS2Hor} we present a large class of AdS$_2$ vacua of eleven-dimensional supergravity which can be viewed as near horizon geometries of black holes constructed out of wrapped M2-branes. We conclude with some comments and a discussion on interesting avenues for future research in Section \ref{sec:discussion}. In the three appendices we collect various technical details used in the main text. In Appendix \ref{app:fieldIIA} we provide some details on the calculation of the topologically twisted index for three-dimensional $\mathcal{N}=2$ SCFTs with massive IIA holographic duals. Appendices \ref{sec:MIIA} and \ref{Appendix11D} contain details on the construction of our massive IIA and eleven-dimensional supergravity solutions. 

\textit{Note added:} While we were preparing this manuscript the papers \cite{Guarino:2017pkw,Guarino:2017eag} appeared. In them the authors find a supersymmetric AdS$_4$ black hole solution, with a near horizon AdS$_2\times \Sigma_\fg$ geometry, in the four-dimensional maximal $ISO(7)$  gauged supergravity. We believe that upon an uplift to massive IIA supergravity this black hole is the same as the universal black hole discussed in Section \ref{sec:upMIIA} below. Soon after this paper appeared on the arXiv other supersymmetric AdS$_4$ black hole solutions in massive type IIA supergravity were studied in \cite{Hosseini:2017fjo,Benini:2017oxt} and their microstates were counted using the topologically twisted index.

%%%%%%%%%%%%%%%%%%%%%%%%%%%%%%%%%%%%%
\section{A universal twist of 3d $\mathcal{N}=2$ SCFTs}
\label{sec:fieldtheory}
%%%%%%%%%%%%%%%%%%%%%%%%%%%%%%%%%%%%%

We consider   $\mathcal{N}=2$ superconformal field theories compactified on a Riemann surface $\Sigma_\fg$ of genus $\fg$ with a topological twist. This is implemented by turning on a non-trivial background for the R-symmetry of the theory. More precisely, there is a magnetic flux on $\Sigma_\fg$ for the  background gauge field $A_R$ coupled to the R-symmetry current such that  $\int_{\Sigma_\fg} d A_R=2\pi (1- \fg)$. This condition ensures that the  R-symmetry background field precisely cancels the spin connection on $\Sigma_\fg$ and the resulting theory preserves two real supercharges. In general, since  the R-symmetry can mix with global symmetries, the choice of $A_R$  is not unique and there is a family of different twists parametrized by the freedom to turn on magnetic fluxes along the Cartan subgroup of the continuous global flavor symmetry group.\footnote{By a flavor symmetry here we mean any global symmetry of the $\mathcal{N}=2$ SCFT which is not the superconformal $U(1)$ R-symmetry. Later on we will make a distinction between different types of non-R symmetries.} On the other hand, in a superconformal field theory there is an exact R-symmetry that is singled out by the superconformal algebra and can be fixed uniquely by performing $F$-maximization \cite{Jafferis:2010un,Closset:2012vg}. Following the terminology in \cite{Benini:2015bwz,BC}, we refer to a partial topological twist of the three-dimensional theory on $\Sigma_\fg$  along the exact superconformal R-symmetry as a \textit{universal twist}.

The degeneracy of ground states of the compactified theory after this partial topological twist can be extracted from the topologically twisted index $Z$, which is defined as  the twisted supersymmetric partition function on $\Sigma_\fg\times S^1$ \cite{Benini:2015noa,Benini:2016hjo,Closset:2016arn}.  Let us briefly summarize some of the salient features of the topologically twisted index.

The index depends on a set of integer magnetic fluxes $\mathfrak{n}= \frac{1}{2\pi}\int_{\Sigma_\fg} F^{{\rm flav}}$  for the Cartan generators of the flavor symmetry group,
parameterizing the inequivalent twists. A convenient parameterization for the fluxes is the following. We can assign
a magnetic flux $\mathfrak{n}_I$  to each chiral field $\Phi_I$ in the theory with the constraint that, for each term $W_a$ in the superpotential,\footnote{In this paper we restrict to three-dimensional $\mathcal{N}=2$ SCFTs which admit a Lagrangian description, however we suspect that the universal twist relation \eqref{entropyfree} below is valid more generally.}
\begin{equation}\label{fluxes} 
\sum_{I\in W_a} \fn_I = 2(1-\fg) \, .
\end{equation}
Since  the superpotential has R-charge $2$, this condition ensures that $\int_{\Sigma_\fg} d A_R=2\pi (1-\fg)$ and supersymmetry is preserved. The Dirac quantization condition further restricts the flux parameters $\fn_I$ to be
integer.  

The index also depends on a set of complex fugacities $y$  for the flavor symmetries. We can again assign a complex number $y_I$   to each chiral field $\Phi_I$ in the theory with the constraint that, for each   term $W_a$ in the superpotential,
\begin{equation}\label{fugacities} 
\prod_{I\in W_a} y_I = 1 \, .
\end{equation}
It will be important also to consider the complexified chemical potentials $\Delta_I$, defined by $y_I \equiv e^{i \Delta_I}$. Notice that  the chemical potentials are  periodic, $\Delta_I\sim \Delta_I +2 \pi$.   Therefore (\ref{fugacities}) becomes
 \begin{equation}\label{chemical0}  
 \sum_{I\in W_a} \Delta_I \in  2 \pi \, \mathbb{Z} \,.
 \end{equation}  
Using this periodicity we can always choose $0\le {\rm Re} \, \Delta_I \le 2\pi$. With this at hand, (\ref{chemical0}) implies that $ \sum_{I\in W_a} \Delta_I \ne 0$ unless all $\Delta_I=0$. This will be important in the discussion below.

The topologically twisted index can be evaluated by localization and reduced to a matrix model  \cite{Benini:2015noa}. The large $N$ limit of the matrix model has been analyzed in \cite{Benini:2015eyy} for the ABJM theory \cite{Aharony:2008ug} and generalized to other classes of quiver gauge theories with M-theory or massive type IIA duals in \cite{Hosseini:2016tor,Hosseini:2016ume}.
The results of this analysis is surprisingly simple. One finds a consistent large $N$ solution of the matrix model only  when 
 \begin{equation}\label{chemical}  
 \sum_{I\in W_a} \Delta_I = 2 \pi \;,
 \end{equation}  
 for each term $W_a$ in the superpotential. Under this condition,  the logarithm of the topologically twisted index is given by\footnote{The formula appears in \cite{Hosseini:2016tor} only for $\Sigma_\fg=S^2$. The generalization to arbitrary $\fg$ is straightforward and discussed in details for ABJM in \cite{Benini:2016hjo}: the general rule is simply $\log Z_{\fg}= (1-\fg) \log Z_{\fg=0} (\fn_I/(1-\fg))\,$.} 
 \begin{equation}\label{index}
 \log Z (\Delta_I,\fn_I) =  (1-\fg) i  \left ( \frac{2}{\pi} \cV (\Delta_I) + \sum_I \left [ \( \frac{\fn_I}{1-\fg} - \frac{\Delta_I}{\pi}  \) \frac{\partial \cV}{\partial\Delta_I} \right ] \right )\;,
\end{equation}
where the function $\cV$ was called Bethe potential in  \cite{Benini:2015noa} and is the Yang-Yang function of an associated integrable system \cite{Nekrasov:2014xaa,Closset:2016arn}.\footnote{What we call $ \cV (\Delta_I)$ here is the extremal value  with respect to the eigenvalues $u_i$  of the Bethe functional  $\cV [\Delta_I, u_i]$, defined in \cite{Benini:2015eyy,Hosseini:2016tor,Hosseini:2016ume}.}  Quite remarkably, in the large $N$ limit, the Bethe potential $\cV$ is related to the free energy on $S^3$ of the three-dimensional $\mathcal{N}=2$ theory  \cite{Hosseini:2016tor} through the simple identity
\begin{equation}
\label{indexS3}
 - \frac{2 i}{\pi} \cV(\Delta_I) = F_{S^3} \left (\frac{\Delta_I}{\pi}\right ) \, . 
 \end{equation}
This relation might look puzzling at first sight and deserves some comments. Recall that the free energy  $F_{S^3}$ is a function of a set of trial R-charges that parameterize a family of supersymmetric Lagrangians on $S^3$ \cite{Jafferis:2010un,Hama:2010av}. The importance of this functional is that its extremization gives the exact R-charges of the theory \cite{Jafferis:2010un}.  In \eqref{indexS3},   $\cV$ is a function of chemical potentials for the flavor symmetries while $F_{S^3}$ is a function of R-charges.  However, although the $\Delta_I$ parameterize flavor symmetries in the three-dimensional theory on $\Sigma_\fg\times S^1$, the relation \eqref{chemical}
ensures that $\Delta_I/\pi$ can be consistently identified with a set of R-charges for the theory on $S^3$. 

There is a subtlety that arises when computing the topologically twisted index or the three-sphere free energy. In three-dimensional $\mathcal{N}=2$ SCFTs there are finite counterterms which affect the imaginary part of the complex function  $F_{S^3}$. These are given by Chern-Simons terms with purely imaginary coefficients for the background gauge fields that couple to conserved currents, see \cite{Closset:2012vg,Closset:2012vp} for a detailed discussion. Moreover, the imaginary part of $\log Z$ is only defined modulo $2\pi$ and is effectively $O(1)$ in the large $N$ limit. The upshot of this discussion is that the physically unambiguous quantity in the large $N$ limit  are the real parts of the topologically twisted index and the free energy on $S^3$. 

There is an additional important point in the story.  To obtain the degeneracy of vacua of the compactified theory  for a given choice of the flux parameters, $\fn_I$, one has to extremize the function $Z (\Delta_I,\fn_I)$ with respect to the fugacities $\Delta_I$ \cite{Benini:2015eyy}. This prescription is analogous to the extremization principles that exist in four \cite{Intriligator:2003jj}, three \cite{Jafferis:2010un}, and two-dimensional \cite{Benini:2012cz,Benini:2013cda} SCFTs with an Abelian R-symmetry.
 
After this short introduction to the topologically twisted index we are ready to discuss  the universal twist. This is obtained by choosing fluxes $\fn_I$ proportional to the exact UV R-charges $\bar \Delta_I$ and these, as we already mentioned,  can be found  by extremizing $F_{S^3}$ \cite{Jafferis:2010un}. From the identification (\ref{indexS3}) it is clear that the Bethe potential $\cV$ is also extremized at the values $\bar \Delta_I$. Given the normalizations (\ref{fluxes}) and (\ref{chemical}),  we find that the universal twist is determined by
 \begin{equation}
 \label{ident} 
\frac{\bar \fn_I}{1-\fg} =  \frac{\bar \Delta_I}{\pi} \, .
\end{equation}
It is easy to see that, for this choice of fluxes,  $ \log Z$ in (\ref{index})  is also extremized at $\Delta_I=\bar\Delta_I$: 
\begin{equation}
\frac{\partial  \log Z}{\partial \Delta_I} \Big |_{\bar \Delta_I} =0 \, ,
\end{equation}
since $\partial \cV (\bar \Delta_I) =0$.\footnote{We note in passing that if one imposes $\frac{ \fn_I}{1-\fg} =  \frac{\Delta_I}{\pi}$ before extremizing $F_{S^3}$ then the second term in \eqref{index} vanishes. Thus, after using \eqref{indexS3} we find that the extremization of $F_{S^3}$ is the same as the extremization of the topologically twisted index.} The  value at the extremum reads 
 \begin{equation}\label{entropyfree}
 \log Z (\bar \Delta_I,\bar\fn_I) =  (\fg-1) F_{S^3} \left (\frac{\bar \Delta_I}{\pi}\right ) \, ,
 \end{equation}
where we made use of \eqref{indexS3} and \eqref{ident}. 

Equation \eqref{entropyfree} amounts to a universal relation between the value of the index of the universal twist of a 3d $\mathcal{N}=2$ SCFT and the free energy on $S^3$ of the same superconformal theory in the planar limit. For theories with a weakly coupled holographic dual this identity should translate into a universal relation between the entropy of some universal black hole solution and the AdS$_4$ supersymmetric free energy. As we discuss in detail in section~\ref{sec:A simple 4d black hole} this expectation indeed bears out for the case of hyperbolic Riemann surfaces, i.e. for $\fg>1$.

It is worth pointing out that the universal relation in \eqref{entropyfree}  is the three-dimensional analog of the universal relation among central charges established in \cite{Benini:2015bwz} for twisted compactifications of four-dimensional $\cN=1$ SCFTs on $\mathbb{R}^2\times \Sigma_{\fg}$.\footnote{Similar relations can be established for SCFTs with a continuous R-symmetry in various dimensions and with different amount of supersymmetry \cite{BC}.} A notable difference is that for four-dimensional SCFTs the universal relation can be established at finite $N$. It would be most interesting to study subleading, i.e. non-planar, corrections to the universal relation in \eqref{entropyfree}.\footnote{It is interesting to note that similar relations between $F_{S^{3}}$ and partition functions on $S^{1}\times \Sigma_{\fg}$ with line operator insertions were discussed in \cite{Closset:2017zgf}.}

%%%%%%%%%%%%%%%%%%%%%%%%%%%%%%%%%%%%%%
\subsection{M-theory and massive type IIA models} 
%%%%%%%%%%%%%%%%%%%%%%%%%%%%%%%%%%%%%%

The derivation of (\ref{entropyfree})  is based on  the large $N$ identities (\ref{index}) and (\ref{indexS3}), which in turn can be established for a large class of Yang-Mills-Chern-Simons theories with fundamental and bi-fundamental chiral fields with M-theory or massive type IIA duals. We now  discuss the class of theories for which these identities are valid.  

Consider first superconformal theories dual to M-theory on AdS$_4\times$ SE$_7$, where SE$_7$ is a Sasaki-Einstein manifold. Many quivers describing such theories have been proposed in the literature. Most of them  are obtained by dimensionally reducing  a ``parent'' four-dimensional quiver gauge theory with bi-fundamentals and adjoints with an AdS$_5\times$ SE$_5$ dual, and then  adding Chern-Simons terms and flavoring with fundamentals.  In such theories, the sum of all Chern-Simons levels is zero, $\sum_a k_a=0$. Holography predicts that the $S^3$ free energy and the twisted index of such theories  scales as $N^{3/2}$ for $N\gg k_a$.  The  large $N$ behavior of the $S^3$ free energy has been computed in   \cite{Jafferis:2011zi} and successfully compared with the holographic predictions only for a particular class of quivers. In particular, for the method in  \cite{Jafferis:2011zi}  to work, the bi-fundamental fields must transform in a real representation of the gauge group and the total number of fundamentals must be equal to the total number of anti-fundamentals.  It turns out that, under the same conditions, the topologically twisted index scales like $N^{3/2}$ and the identities (\ref{index}) and (\ref{indexS3}) are valid \cite{Hosseini:2016tor}. 
This particular class of quivers include all the vector-like examples in \cite{Hanany:2008fj, Hanany:2008cd, Martelli:2008si} and many of the flavored theories in  \cite{Gaiotto:2009tk,Benini:2009qs}. In particular, the latter  includes the dual of AdS$_4\times Q^{1,1,1}/\mathbb{Z}_k$ and AdS$_4\times N^{0,1,0}/\mathbb{Z}_k$.  The conditions are also satisfied for the $\mathcal{N}=3$ necklace and  $\widehat D$- and $\widehat E$-type quivers  \cite{Jafferis:2008qz,Herzog:2010hf,Gulotta:2011vp,Crichigno:2012sk,Crichigno:2017rqg}, as well as the quiver for the non-toric manifold $V^{5,2}$ discussed in \cite{Martelli:2009ga}. The evaluation of the index in the large $N$ limit for most of these examples is given in \cite{Hosseini:2016ume}, where the identities (\ref{index}) and (\ref{indexS3})  are also verified by explicit computation. For the ``chiral'' theories discussed in \cite{Hanany:2008fj, Hanany:2008cd, Martelli:2008si}, on the other hand,  it is not known how to properly take the large $N$ limit in the matrix model to obtain the correct scaling predicted by holography. This applies both for the topologically twisted index and for the $S^3$ partition function. This class of quivers includes interesting models, like the quiver for $M^{1,1,1}$ proposed in \cite{Martelli:2008si} and further studied in \cite{Benini:2011cma}.

Consider now superconformal theories dual to warped AdS$_4\times Y_6$  flux vacua of massive type IIA. A well-known example is the $\mathcal{N}=2$ $U(N)$  gauge theory with three adjoint
multiplets  and a Chern-Simons coupling $k$ described in \cite{Guarino:2015jca}.  It corresponds to 
an internal manifold $Y_6$ with the topology of $S^6$. This has   been generalised in  \cite{Fluder:2015eoa} to the case where $Y_6$ is  an $S^2$ fibration over  a general K\"ahler-Einstein manifold KE$_4$.
The dual field theory  is obtained by considering the four-dimensional theory dual to AdS$_5\times$ SE$_5$, where SE$_5$ is the five-dimensional Sasaki-Einstein
with local base KE$_4$, reducing it to three dimensions and adding a Chern-Simons term with level $k$ for all gauge groups.\footnote{The original example in \cite{Guarino:2015jca} has KE$_4=\Bbb{CP}_2$, SE$_5=S^5$,  and the superconformal theory is obtained by reducing $\mathcal{N}=4$ SYM to three-dimensions and adding a Chern-Simons coupling with level $k$. The solution in \cite{Fluder:2015eoa}
is obtained by replacing $\Bbb{CP}_2$ with more general KE$_4$ manifolds.} The large $N$ limit at fixed $k$ of the $S^3$ free energy has been computed in \cite{Guarino:2015jca} and  \cite{Fluder:2015eoa} and shown to scale as $N^{5/3}$, as predicted by holography. The large $N$ limit of the topologically twisted index
has been computed in \cite{Hosseini:2016tor}. The  identities (\ref{index}) and (\ref{indexS3}) also hold for massive type IIA. The explicit derivation was not reported in  \cite{Hosseini:2016tor}
and is given in Appendix \ref{app:fieldIIA}. Notice that for massive type IIA quivers there is no need to restrict to vector-like models.  

The  discussion above shows that there is a large number of three-dimensional superconformal theories with M-theory or massive type IIA  duals for which relation (\ref{entropyfree}) formally holds.
However, it is important to notice that not all of them really admit a universal twist since we  need to restrict ourselves to $\mathcal{N}=2$ SCFTs with rational  R-charges. Indeed, since $\fn_I$ and $\fg$ are integers, the relation in (\ref{ident}) implies that the  exact R-charge of the fields must be rational.  This slightly restricts the class of theories where we can perform the universal twist. However, we can still find infinitely many models for which the R-charges are rational and the universal twist exists. Since $\mathcal{N}=3$ theories necessarily have rational R-charges this applies to the  $\mathcal{N}=3$ necklace quivers  \cite{Jafferis:2008qz,Herzog:2010hf} as well as the  $\widehat D$- and $\widehat E$-type quivers \cite{Gulotta:2011vp,Crichigno:2012sk,Crichigno:2017rqg}, and the $\mathcal{N}=3$ quiver for $N^{0,1,0}$ \cite{Gaiotto:2009tk,Imamura:2011uj,Cheon:2011th}. In addition, one can check that the $\mathcal{N}=2$ quivers for $Q^{1,1,1}$ \cite{Benini:2009qs} and $V^{5,2}$  \cite{Martelli:2009ga} in M-theory have rational R-charges. The same holds for the theory in \cite{Guarino:2015jca}
and some of its generalizations in massive type IIA.\footnote{The R-charges for M-theory vacua can be computed using volume minimization \cite{Martelli:2006yb} and for massive type IIA
by $a$-maximization \cite{Intriligator:2003jj}. The result is generically irrational. However, we can easily find special classes of SE$_7$ or SE$_5$ where the result is rational.}

%%%%%%%%%%%%%%%%%%%%%%%%%%%%%%%%%%%%%
\section{A simple 4d black hole}
\label{sec:A simple 4d black hole}
%%%%%%%%%%%%%%%%%%%%%%%%%%%%%%%%%%%%%

Here we provide the holographic description of the universal twisted compactification of  3d  $\cN=2$ SCFTs discussed in the previous section. As we shall show, this corresponds to the supersymmetric magnetically charged AdS$_{4}$ black hole of \cite{Romans:1991nq,Caldarelli:1998hg}, thus providing the appropriate field theory interpretation of this solution. We also review the known uplift of this solution to M-theory and provide a new uplift to massive IIA supergravity. 

The appropriate supergravity is 4d minimal $\cN = 2$ gauged supergravity \cite{Freedman:1976aw}, with eight supercharges and bosonic content the graviton and an $SO(2)$ gauge field, dual to the stress energy tensor and R-symmetry current, respectively. The bosonic action reads\footnote{Here we follow the conventions in  \cite{Benini:2015eyy} and truncate the  ${\cal N}=8$ supergravity to the minimal one by setting $L_a=1$ ($\phi_{ij}=0$), $A_a=A$, and set the coupling constant $g=1/\sqrt{2}$ which in turn amounts to setting the radius of the AdS$_{4}$ vacuum to one.}  
\begin{equation}\label{eq:4daction}
I= \frac{1}{16\pi G_N^{(4)}} \int d^4x \sqrt{-g} \(R +6 -\frac14 F^{2}\) \,,
\end{equation}
with $G_N^{(4)}$ the 4d Newton constant. We have chosen the cosmological constant such that the AdS$_4$ vacuum of the theory has  $R_{\text{AdS}_{4}}=-12$ and $L_{\text{AdS}_{4}}=1$. The magnetically charged black hole solution of \cite{Romans:1991nq,Caldarelli:1998hg} preserves two supercharges and is given by\footnote{A generalization to include rotation while maintaining supersymmetry was also found in these references.}
\eqss{ \label{univ}
ds_4^2 &=   -\left ( \rho -\frac{1}{2\rho}\right )^2 dt^2 + \left ( \rho -\frac{1}{2\rho}\right )^{-2} d\rho^2 + \rho^2\, ds_{\Bbb H^2}^2 \,,  \\
F&= \frac{dx_1 \wedge dx_2}{x_2^2} \,,
}
where $ds_{\Bbb H^2}^2 = \frac{1}{x_2^2}\left(dx_1^2+dx_2^2\right) $ is the local constant-curvature metric on a Riemann surface $\Sigma_\fg$ of genus $\fg>1$,\footnote{As discussed in \cite{Caldarelli:1998hg}, there is no supersymmetric static black hole solution with $\fg=0,1$.} normalized such that $R_{\Bbb H^2}=-2$. Using the Gauss-Bonnet theorem the volume of the Riemann surface is then ${\rm vol}(\Sigma_\fg)=4 \pi(\fg-1)$.  Dirac quantization of the flux requires $\frac{1}{2\pi}\int_{\Sigma_{\fg}}F\in \Bbb Z$, which holds for any genus $\fg$. We note the solution has a fixed magnetic charge, set by supersymmetry. The entropy of this extremal black hole is given in terms of the horizon area by the standard Bekenstein-Hawking formula
\begin{equation}\label{eq:SBH}
S_\text{BH} = \frac{\text{Area}}{4 G_N^{(4)}} = \frac{(\fg-1)\,\pi}{2 G_N^{(4)}} \, .
\end{equation}
As will soon become clear the large $N$ limit of the topologically twisted index reproduces exactly this entropy. However, there is a slight subtlety in this story. According to the standard holographic dictionary the logarithm of the partition function of the CFT (in the appropriate large $N$ limit) should correspond to a properly regularized on-shell action of the dual gravitational background, rather than the black hole entropy. In the next section we clarify this relation,  showing that the black hole entropy in \eqref{eq:SBH} is indeed closely related to the on-shell action and thus to the topologically twisted index.

%%%%%%%%%%%%%%%%%%%%%%
\subsection{On-shell action}
%%%%%%%%%%%%%%%%%%%%%%

We are interested in calculating the value of the  Euclideanized action \eqref{eq:4daction}, evaluated on-shell for the solution \eqref{univ}. This action is divergent unless properly regularized by counterterms, following the standard holographic renormalization prescription, which we carry out explicitly next. As we show, this is intimately related to the entropy of the black hole. The regularized action we consider is given by:\footnote{See for example Section 4.4 of \cite{Martelli:2012sz} as a reference for the counterterms. With respect to their normalizations we have $F^\text{there} = F^\text{here}/2$.}
\begin{equation}\label{Ireg}
\begin{split}
I_\text{Eucl} =\,& I_\text{Einst+Max} + I_\text{ct+bdry} \,, \\
I_\text{Einst+Max} =\,&-\frac{1}{16\pi G_N^{(4)}} \int d^4 x \sqrt{g} \left(R+6-\frac{1}{4} F^2 \right) \\
I_\text{ct+bdry} = \,&\frac{1}{8\pi G_N^{(4)}} \int d^3x \sqrt{\gamma} \left(2+\frac{1}{2}R(\gamma)-K\right) \, ,
\end{split}
\end{equation}
where $\gamma$ is the induced metric on the boundary, defined by the radial cutoff $\rho=r$, and $K$ is the trace of the extrinsic curvature of this boundary metric. Taking $r\rightarrow \infty$ leads to a divergence in $I_\text{Einst+Max}$, which is cancelled by $I_\text{ct+bdry}$, where we have collected the appropriate counterterms as well as boundary terms necessary for a well-defined variational principle. An important additional subtlety arises, however, in the explicit evaluation of $I_\text{Eucl}$ for the extremal solution \eqref{univ}. Indeed, it is easy to see that this integral is naively not well defined,   as the integrand of $I_\text{Eucl}$ for this solution vanishes, while the integration over Euclidean time $\int_{0}^{\infty} d\tau$ leads to infinity as a consequence of the solution being  extremal and thus $T=0$. To obtain the correct finite result we thus consider a non-extremal deformation of the solution, compute $I_\text{Eucl}$ for the non-extremal solution and take the extremal limit at the end.

There are two non-extremal deformations. One amounts to allowing for a generic  magnetic charge $Q$ under the graviphoton, and a second to  adding a mass $\eta$. This non-extremal generalization was discussed in \cite{Romans:1991nq,Caldarelli:1998hg} and the solution reads 
\eqss{\label{eq:nonextr}
ds^2 &= -V(\rho) \, dt^2 +\frac{1}{V(\rho)} \, d\rho^2 + \rho^2 \, ds_{\Bbb H^2}^2  \, ,\\
F &= 2Q \,\frac{dx_1 \wedge dx_2}{x_{2}^{2}}\,,
}
where 
\begin{equation}
V(\rho) = -1-\frac{2\eta}{\rho} + \frac{Q^2}{\rho^2}+\rho^2 \, .
\end{equation}
The extremal solution is recovered for $Q\rightarrow 1/2$ and $\eta\rightarrow 0$. The horizon radius $\rho_0$ is obtained by solving the quartic equation $V(\rho_0)=0$. The temperature $T$ of the black hole can be obtained by requiring the Euclidean metric to be free of conical singularities, which gives 
\begin{equation}
T = \frac{\left|V'(\rho_0)\right|}{4\pi} = \frac{1}{2\pi \rho_0} \left|\rho_0^2 + \frac{\eta}{\rho_0}-\frac{Q^2}{\rho_0^2}\right|\, .
\end{equation}
In the extremal limit $\rho_0 \rightarrow 1/\sqrt{2}$ and $T\rightarrow0$.
Evaluating the on-shell action \eqref{Ireg} for general values of $Q$ and $\eta$ we find
\begin{equation}\label{actionnonext}
I_\text{Eucl} = \frac{(\fg-1)}{2 G_N^{(4)}}\frac{\beta}{\rho_0} \left(Q^2-\rho_0^4+\eta \rho_0\right) \, ,
\end{equation}
where $\beta=1/T$ is the Euclidean time periodicity. Taking the extremal limit of this expression gives the finite answer
\begin{equation}
I_\text{extr}= - \frac{ (\fg-1)\,\pi}{2 G_N^{(4)}} + \mathcal{O}\left(\left(Q-1/2\right)^{1/2}\right)+\mathcal{O}\left(\eta^{1/2}\right) \, .
\end{equation}
Comparing this to the entropy \eqref{eq:SBH} of the extremal black hole, we thus have
\begin{equation}\label{eq:ISBH}
 I_\text{extr}=-S_\text{BH}  \, .
\end{equation}
On the other hand, the holographic dictionary relates the gravitational on-shell action to the partition function of the dual CFT as $I=-\log Z$, leading to the satisfying expression 
\begin{equation}\label{eq:ident}
S_\text{BH}=\log Z  \,.
\end{equation}
We have thus shown that the black hole entropy can be identified with the topologically twisted index of the dual CFT  to leading order in $N$. This relation was argued to hold more generally for a class of black holes in non-minimal gauged supergravity in \cite{Benini:2015eyy}; it would be interesting to establish this explicitly by generalizing the computation above to this larger class of black holes. It would also be of interest to study these relations to subleading orders in $N$. 

Finally, let us recall that the renormalized on-shell action for Euclidean AdS$_4$ with an $S^3$ boundary is given by  \cite{Emparan:1999pm}\footnote{With slight abuse of notation we define the CFT free energy as $F=-\log Z=I$.}
\begin{equation}
F_{S^3} = \frac{\pi }{2 G_N^{(4)}} \, ,
\end{equation}
which using \eqref{eq:SBH} leads to the relation
\begin{equation}\label{eq:SBHFS3}
S_\text{BH} = (\fg-1) \, F_{S^3} \, .
\end{equation}
With the identification \eqref{eq:ident} one recognizes this result as the holographic analog of the universal field theory relation \eqref{entropyfree}. Thus, the result \eqref{eq:SBHFS3} provides strong evidence that the AdS$_{4}$ black hole  \eqref{univ} describes the RG flow from 3d $\cN=2$ SCFTs with a universal topological twist on $\Sigma_{\fg}$ to superconformal quantum mechanical theories with two supercharges.

We emphasize that although the black hole solution \eqref{univ} is derived as a supersymmetric solution of minimal gauged supergravity, it is also a solution to non-minimal gauged supergravity, with the additional vector and hyper multiplet bosonic fields set to zero.\footnote{This can be checked, for instance, by  setting $\mathfrak n_{a}=\frac{\kappa}{2}\,, \vec \phi=0\,, L_{a}=1$ in the BPS equations (A.27) in \cite{Benini:2015eyy}.} The fact that ``freezing'' the vector multiplets to zero is consistent corresponds to the property of universality of the topological twist \eqref{ident}. In other words, as the vector multiplet scalars in the gauged supergravity are not sourced along the flow from AdS$_4$ to AdS$_2$ realized by the black hole  \eqref{univ}, there is no mixing between R-symmetry and flavor symmetry along the flow and the R-symmetry in the IR coincides with the one in the UV. This is consistent with the field theory discussion of Section~\ref{sec:fieldtheory}.  

%%%%%%%%%%%%%%%%%%%%%%%%%%%%%%%%%%%%%
\subsection{Uplift to M-theory}
\label{sec:MTheoryUplift}
%%%%%%%%%%%%%%%%%%%%%%%%%%%%%%%%%%%%%

Here we review the uplift of the universal solution in \eqref{univ} to eleven-dimensional supergravity, which was carried out in  \cite{Gauntlett:2007ma}. The metric and four-form read\footnote{Compared to Equation (2.3) in \cite{Gauntlett:2007ma} we have introduced an overall scale $L$.}
\eqss{\label{redAnsatz11d}
ds_{11}^2 &= L^2\left (\frac{1}{4}\, ds_4^2 + ds_7^2\right )\,, \\
ds_7^2 &= \(d\psi + \sigma + \frac{1}{4}\, A\)^2 + ds_{\rm{KE}}^2\,,  \\
G_4 &= L^3\left (  \frac{3}{8}\, {\rm vol}_4 -\frac{1}{4}  *_4 F \wedge J \right )\,,
}
where $ds_7|_{A=0}$ is a seven-dimensional  Sasaki-Einstein  with   $R^{\rm{SE}_{7}}_{ij}= 6 g^{\rm{SE}_{7}}_{ij}$, $ds_{\rm{ KE}}$ is a six-dimensional K\"ahler-Einstein space with $R^{\rm{KE}}_{ij}= 8g^{\rm{KE}}_{ij}$ and K\"ahler form $J$, with $d\sigma=2 J$, $F=dA=\text{vol}_{\mathbb{H}^2}$, $ds_{4}^{2}$ is given in \eqref{univ}, and $*_4$ is with respect to this four-dimensional metric. This solution of eleven-dimensional supergravity can be interpreted as the backreaction of $N$ M2-branes wrapping a supersymmetric cycle in a Calabi-Yau five-fold. This point of view was discussed in \cite{Gauntlett:2001qs}, following the ideas in \cite{Maldacena:2000mw}, for the case where the Sasaki-Einstein manifold is $S^7$ and thus the six-dimensional K\"ahler-Einstein space is $\mathbb{CP}^3$.

In order for this local supergravity solution to extend to a well defined M-theory background, some of its parameters should be properly quantized. The quantization condition on $G_{4}$ reads
\equ{
N=\frac{1}{(2\pi \ell_{11})^{6}}\int_{Y_{7}}\ast_{11}\, G_{4}\,,
}
where $N$ is an integer determining the number of M2-branes and $\ell_{11}$ is the Planck length in eleven dimensions. This translates into the quantization of the AdS$_4$ length scale in terms of the Planck length,\footnote{Upon a reduction to 10d type IIA supergravity the 11d Planck length is related to the string length and coupling constant via the equation $\ell_{ 11}=l_{s} g_{s}^{1/3}$.}
\equ{\label{quant L ads2}
L=\pi \ell_{11}\,\(\frac{32N}{3\vol(Y_{7})}\)^{1/6}\,.
}
The four-dimensional Newton constant is given by
\equ{\label{GN4}
\dfrac{1}{G_{N}^{(4)}}=\dfrac{L^{7}\, \vol(Y_{7})}{ G_{N}^{(11)}}\,, \qquad 16\pi G_{N}^{(11)}= (2\pi)^{8}\ell_{11}^{9}\,.
}

%%%%%%%%%%%%%%%%%
\subsubsection{Entropy}
%%%%%%%%%%%%%%%%%

Using  \eqref{quant L ads2} and \eqref{GN4} we compute the properly normalized horizon area,  obtaining the black hole entropy:
\equ{\label{univM2}
S_{\rm{BH}}=(\fg-1)\,\sqrt{\frac{2\pi^{6}}{27\vol(Y_{7})}}\,N^{3/2}\,.
}
This is consistent with \eqref{eq:SBHFS3} and the well-known expression for the holographic free energy on $S^3$ \cite{Herzog:2010hf,Drukker:2010nc}
\begin{equation}
F_{S^{3}}=\sqrt{\frac{2\pi^{6}}{27\vol(Y_{7})}}\,\,N^{3/2}\;.
\end{equation}
It is instructive to unpack this equation for a couple of examples of well-known three-dimensional SCFTs. For the special case of the ABJM theory, i.e., $Y_{7}=S^{7}/\Bbb Z_{k}$,  the free energy on $S^{3}$ is given by  $F_{S^{3}}=\frac{\pi\sqrt 2}{3}k^{1/2}N^{3/2}$ \cite{Drukker:2010nc} and thus 
\equ{
S_{\rm BH}^{\rm ABJM}= (\fg-1)\, \frac{\pi\sqrt 2}{3}k^{1/2}N^{3/2}\,.
}
This result is correctly reproduced by the topologically twisted index for the ABJM theory. This follows from the general discussion  in Section \ref{sec:fieldtheory}, which is model independent.
It is nonetheless instructive to see explicitly how this works.  The topologically twisted index for ABJM is given by \cite{Benini:2015eyy,Benini:2016rke} 
\equ{\label{topindexABJM} 
\log Z^{\rm ABJM}= -\frac{\sqrt 2}{3}k^{1/2}N^{3/2} \sqrt{\Delta_1\Delta_2\Delta_3\Delta_4} \, \sum_{I=1}^4 \frac{\fn_I}{\Delta_I} \, ,
} 
where $\Delta_I,\fn_I$ are, respectively, the chemical potentials and fluxes associated to the four chiral bi-fundamental fields of ABJM, subject to the constraints $\sum_{I=1}^4 \Delta_I=2\pi$ and $\sum_I \fn_I= 2(1-\fg)$.\footnote{See  formula (28) in \cite{Benini:2016rke}, where the notations are: $u_I=\Delta_I$ and $\mathfrak{p}_I=-\fn_I$ and the case $k=1$ is considered. For $k>1$ there could be ambiguities related to other saddle points, but we still expect (\ref{topindexABJM}) to hold.}  The exact R-symmetry corresponds to $\Delta_I=\pi/2$ and the universal twist to $\fn_I=(1-\fg)/2$. We see that $\fg$ must be odd. We obtain 
$\log Z^{\rm ABJM} = S_{\rm BH}^{\rm ABJM}$, as expected.

As another example, we can consider the Sasaki-Einstein manifold $Y_{7}=Q^{1,1,1}/\Bbb Z_{k}$ whose dual SCFT was discussed in  \cite{Benini:2009qs}. The free energy on $S^{3}$ is given by  $F_{S^{3}}=\frac{4\pi}{3\sqrt{3}}k^{1/2}N^{3/2}$ \cite{Jafferis:2011zi}. Then, from  \eqref{eq:SBHFS3} we have
\equ{
S_{\rm BH}^{Q^{1,1,1}}= (\fg-1)\,\frac{4\pi}{3\sqrt{3}}k^{1/2}N^{3/2} \, .
}
The dual field theory is a flavored version of ABJM  and the exact R-charge of the fields $A_i$ is $2/3$ and that of $B_i$ is $1/3$ \cite{Jafferis:2011zi}. Quantization of fluxes then requires $\fg-1$ to be an integer multiple of $3$.  The topologically twisted index was computed in  \cite{Hosseini:2016ume} and it is easy to check that  $\log Z^{ Q^{1,1,1}} = S_{\rm BH}^{Q^{1,1,1}}$, again in agreement with 
the general result  in Section \ref{sec:fieldtheory}.\footnote{The free energy on $S^{3}$ is given by formula (6.15) in \cite{Jafferis:2011zi} and the twisted index for $\fg=0$ by formula (5.47) in \cite{Hosseini:2016ume}. The generalization to general genus $\fg>0$  is obtained by $\log Z_{\fg}= (1-\fg) \log Z_{\fg=0} (\fn_i/(1-\fg))$  \cite{Benini:2016hjo}.   As discussed in \cite{Jafferis:2011zi}, the exact R-charge of
the fields $A_i$ is $\Delta_{A_i}/\pi=2/3$ and that of the fields $B_i$ is $\Delta_{A_i}/\pi=1/3$, while   $\Delta_m$=0.  The  fluxes for the universal twist are then  $n_{A_i}=2(1-\fg)/3$,  $n_{B_i}=(1-\fg)/3$  and $\mathfrak{t}+\mathfrak{\tilde t}=0$. From formula (5.47) in \cite{Hosseini:2016ume} we  find $\log Z^{Q^{1,1,1}} = (\fg-1)\frac{4\pi}{3\sqrt{3}}k^{1/2}N^{3/2}$.}
Combining the results in \cite{Jafferis:2011zi} and  \cite{Hosseini:2016ume} one can check many other examples, all in agreement with the general result (\ref{entropyfree}). 

As discussed in Section \ref{sec:fieldtheory}, the universal twist is not only possible for theories with an M-theory dual, but also for field theories with massive IIA duals with an $N^{5/3}$ scaling of the free energy.  We discuss this next.

%%%%%%%%%%%%%%%%%%%%%%%%%%%%%%%%%%%%%
\subsection{Uplift to massive IIA}
\label{sec:upMIIA}
%%%%%%%%%%%%%%%%%%%%%%%%%%%%%%%%%%%%%

Here we discuss new black hole solutions in massive IIA supergravity, obtained by uplifts of the 4d solution \eqref{univ}. This can be done by using the formulae of \cite{Guarino:2015vca,Varela:2015uca,Guarino:2015qaa}, where the uplift of the SU$(3)$-invariant sector of 4d $\cN=8$  supergravity with ISO$(7)$ gauging is given. The bosonic content of the 4d SU$(3)$-invariant sector is the graviton $e_{\mu}^{a}$, 6 scalars $\varphi, \chi, \phi, a,\zeta, \tilde \zeta$, 2 electric vectors $A^{0},A^{1}$ and their magnetic duals  $\tilde A_{0},\tilde A_{1}$, 3 two-forms $B^{0},B_{1},B_{2}$ and 2 three-forms $C^{0},C^{1}$. One should keep in mind that these are not all independent as, e.g., the field strengths of $C^{0},C^{1}$ can be dualized into functions of the scalar fields, see \cite{Guarino:2015qaa} for more details. 

Since we are interested in solutions that asymptote to the $\mathcal{N}=2$ supersymmetric AdS$_{4}$ vacuum of the theory described in \cite{Guarino:2015jca}, and we expect the scalars to be set to constant values, we set them to the values of the AdS$_4$ vacuum solution, provided in  \eqref{eq:scalarIIA}. The electric potential $A^{1}$ is identified (up to a normalization, which we fix by using the massive IIA equations of motion) with the gauge field $A$ of the minimal four-dimensional theory in \eqref{eq:4daction}.  With this at hand we can use the uplift formulae of \cite{Guarino:2015vca,Varela:2015uca,Guarino:2015qaa}, reproduced in \eqref{eq:uplift1}, to obtain explicit expressions for the bosonic fields of the ten-dimensional massive IIA supergravity, i.e., the metric, the dilaton, $\hat{\phi}$, as well as the two-, three-, and four-form fluxes: $\hat{F}_2$, $\hat{H}_3$, and $\hat{F}_4$, respectively. The metric and dilaton read 
\eqss{\label{flowIIAmetric}
ds^{2}_{10}&=L^{2}(\cos (2\alpha)+3)^{1/2}(\cos(2\alpha)+5)^{1/8}\(ds^{2}_{4}+ds^{2}_{6}\)\,,\\
ds^{2}_{6}&=\frac32 d\alpha^{2}+\frac{6 \sin ^2(\alpha )}{\cos (2 \alpha )+3} \, ds^{2}_{\Bbb{CP}^{2}}+\frac{9 \sin ^2(\alpha )}{\cos (2 \alpha )+5}\,( d\psi + \sigma +  g A^1)^{2}\,,\\
e^{\hat{\phi}} &= 2^{1/4} g^{5/6} m^{-5/6} \, \frac{(\cos (2 \alpha )+5)^{3/4}}{\cos (2 \alpha )+3}\,,
}
where $ds^{2}_{4}$ is the black hole metric \eqref{univ}, $ds^{2}_{\Bbb{CP}^{2}}$ is the standard Einstein metric on ${\mathbb{CP}^2}$ (see \eqref{CP2}), $A^1 = \frac{1}{3g} \frac{dx_1}{x_2} $ is the connection on the Riemann surface, and 
\equ{\label{LmassiveIIA}
L^{2}= 3^{-1} 2^{-5/8}g^{-25/12}m^{1/12}\,.
} 
We note that \eqref{flowIIAmetric} can be obtained from the AdS$_{4}$ vacuum solution of \cite{Guarino:2015jca} by the simple replacement $ds^{2}_{\text{AdS}_{4}}\to ds^{2}_{4}$ and $d\psi\to d\psi+gA^1$.  The massive IIA form fields, however, are not so easily obtained and require more work. The basic point is that in order to determine these one must find a further truncation of the remaining $SU(3)$-invariant fields consistent with the duality transformations of \cite{Guarino:2015qaa} and the equations of motion, which due to the connection on the Riemann surface is non-trivial. We discuss this in detail in Appendix~\ref{sec:AppUplift}. Here we simply present the final answer:
\begin{equation}\label{eq:uplift2A}
\begin{split}
\hat{F}_2 = & \, \left(\frac{m}{g}\right)^{2/3} \Bigg(-\frac{4 \sin ^2(\alpha ) \cos (\alpha )}{g (\cos (2 \alpha )+3) (\cos (2 \alpha )+5)}\, J - \frac{3 \sin (\alpha ) (\cos (2 \alpha )-3) }{g (\cos (2 \alpha )+5)^2}\, \eta \wedge d\alpha \\
& + \frac{3 \cos (\alpha )}{\cos (2 \alpha )+5} H^{1}_{(2)} -\frac{1}{2} \sqrt{3}\cos (\alpha ) \ast_4 H^{1}_{(2)} \Bigg) \, , \\
\hat{H}_3 = & \, \frac{1}{m} \left(\frac{m}{g}\right)^{2/3} \Bigg( \frac{8 \sin ^3(\alpha )}{g (\cos (2 \alpha )+3)^2}\, d\alpha \wedge J + \frac{1}{2} \sqrt{3} \sin (\alpha ) \, d\alpha \wedge \ast_4 H^{1}_{(2)}\Bigg) \, ,\\
\hat{F}_4 =& \, \frac{1}{g^2} \left(\frac{m}{g}\right)^{1/3} \Bigg( \frac{\text{vol}_4}{\sqrt{3}g} +\frac{\sin ^4(\alpha ) (3 \cos (2 \alpha )+7)}{g (\cos (2 \alpha )+3)^2}\, J\wedge J \\
& -\frac{3 \sin ^3(\alpha ) \cos (\alpha ) (\cos (2 \alpha )+9) }{g (\cos (2 \alpha )+3) (\cos (2 \alpha )+5)} \, \eta \wedge d\alpha \wedge J +\frac{3}{4} \sin (\alpha ) \cos (\alpha ) \,  \eta \wedge d\alpha \wedge H^{1}_{(2)} \\
& -\frac{3 \sin ^2(\alpha ) \cos ^2(\alpha )}{\cos (2 \alpha )+3} \, H^{1}_{(2)}\wedge J + \frac{3 \sqrt{3} \sin (2 \alpha ) }{4 (\cos (2 \alpha )+5)} \, \eta \wedge d\alpha \wedge \ast_4 H^{1}_{(2)} -\frac{\sqrt{3} \sin ^2(\alpha )}{\cos (2 \alpha )+3} \ast_4 H^{1}_{(2)} \wedge J \Bigg) \, ,
\end{split}
\end{equation}
where $J=\frac12 d\sigma$ is the K\"ahler form on $\mathbb{CP}^2$, $H^{1}_{(2)}\equiv dA^1$, $\text{vol}_4$ is the volume form of $ds_4$, and $\ast_{4}$ is the hodge dual with respect to $ds_4$. We have explicitly checked that  \eqref{flowIIAmetric} and \eqref{eq:uplift2A} indeed satisfy the equations of motion of massive IIA supergravity.\footnote{These equations can be found in Equation (A.5) of \cite{Guarino:2015vca}.}  At large $\rho$ the solution asymptotes to the $\mathcal{N}=2$ AdS$_4$ solution of \cite{Guarino:2015jca} and thus \eqref{flowIIAmetric} and \eqref{eq:uplift2A} describe the twisted compactification of the corresponding three-dimensional field theory dual, consisting of a  $U(N)$ Chern-Simons theory at level $k$ with three adjoints superfields $\Phi_{i}$. 

Based on our field theory analysis we actually expect \eqref{flowIIAmetric} and \eqref{eq:uplift2A} to be part of a more general class of solutions, where the $\mathbb{CP}^2$  is replaced by a general K\"{a}hler-Einstein base. As an example, we have explicitly checked  that replacing 
\begin{equation}
\begin{split}
ds^2_{\mathbb{CP}^2} &\rightarrow ds^2_{\Bbb{P}^{1}\times \Bbb{P}^{1}} = \frac{1}{6} \left(d\theta_1^2+\sin^2 \theta_1 \, d\phi_1^2 + d\theta_2^2+\sin^2 \theta_2 \, d\phi_2^2 \right) \, , \\
\sigma &\rightarrow \frac{1}{3} \left( \cos \theta_1 \,d\phi_1 + \cos \theta_2 \, d\phi_2 \right) \,,
\end{split}
\end{equation}
in \eqref{flowIIAmetric} (along with suitable replacements of $J$ and $\eta$ in \eqref{eq:uplift2A}) is also a solution of the equations of motion. 

More generally, replacing $\mathbb{CP}^2$ by a generic K\"ahler-Einstein base $\cB$ leads to a large family of new massive IIA black holes, which are  asymptotic to the AdS$_{4}$ solutions described in \cite{Fluder:2015eoa}. As discussed there, the  3d field theory duals are obtained from a 4d parent field theory with an AdS$_{5}\times Y_{5}$ dual, with $\cB$ corresponding to the K\"ahler-Eistein base of the Sasaki-Einstein $Y_{5}$, i.e., 
\equ{ds^{2}_{Y^{5}}=\eta^{2}+ds^{2}_{\cB}
\,.}
Thus, the black hole solutions reported here describe the compactifications of these 3d SCFTs on a Riemann surface, twisted by the exact superconformal R-symmetry. 

To ensure that the local supergravity solution given in \eqref{flowIIAmetric} and \eqref{eq:uplift2A} extends to a well defined string theory background, it should be properly quantized.\footnote{In the case of $ds^{2}_{4}= ds^{2}_{ \rm{AdS}_{4}}$ this was carried out in \cite{Fluder:2015eoa} for a general base $\cB$. The quantization for the black hole solution here is very similar and we provide it here for completeness.} The four-form quantization condition in massive IIA reads\footnote{See for example Equation  (4.11) in \cite{Varela:2015uca}.}
\equ{\label{quantconditionmIIA}
N=-\frac{1}{(2\pi \ell_s)^5}\int_{M_6}\(e^{\frac12 \hat \phi} \ast_4 \hat F_{(4)}+\hat B_{(2)}\wedge d\hat A_{(3)}+\frac16 \, m\, \hat B_{(2)}\wedge \hat B_{(2)}\wedge \hat B_{(2)}\,\),
}
where $N$ is an integer determining the number of D2-branes,  and the potentials $\hat{B}_{(2)}, \hat{A}_{(3)}$ can be found in \cite{Varela:2015uca}. We can evaluate this for the solution in \eqref{flowIIAmetric} and \eqref{eq:uplift2A}  for a general base $\cB$. 
Using the explicit expression for the background fluxes presented in Appendix \ref{sec:AppUplift} we find
\eqs{ \nonumber 
N &= \frac{1}{(2\pi \ell_s)^5} \int_{M_{6}} \frac{\sin ^5(\alpha ) (132 \cos (2 \alpha )+7 \cos (4 \alpha )+213)}{4 g^5 (\cos (2 \alpha )+3)^3} \,d\alpha \wedge J \wedge J \wedge d\psi  \\ \label{eq:N}
&= \frac{1}{(2\pi \ell_s)^5}  \frac{16}{3g^{5}} \, \vol(Y_{5})  \, ,
}
where we integrated over $0\leq\alpha\leq\pi$ and used the identity $\frac12 J\wedge J\wedge d\psi=\dvol (Y_{5})$. Using 
\eqref{LmassiveIIA} this translates into a quantization condition on the asymptotic AdS$_{4}$ and the near-horizon AdS$_{2}$ radii: 
\equ{
L_{\text{AdS}_2}=\frac 12 L_{\text{AdS}_4}=\frac{\pi \ell_{s}n^{1/24}}{2^{55/48}3^{7/24}}\(\frac{N}{\vol(Y_{5})}\)^{5/24}\,,
}
where we have used that asymptotically $ds^{2}_{4}\to ds^{2}_{ \rm{AdS}_{4}}$, while close to the horizon $ds^{2}_{4}\to \frac14 (ds^{2}_{ \rm{AdS}_{2}}+2ds^{2}_{\Sigma_{\fg}})$ and defined $n\equiv2\pi \ell_s m$.
Note that the quantization condition on $L_{\text{AdS}_4}$ coincides with the one obtained in \cite{Fluder:2015eoa}. This is expected  since our solution is asymptotic to the solutions discussed there.  Similarly, the four-dimensional Newton constant is given by\footnote{For a solution of the form 
$ds^{2}_{10}=e^{2\lambda}L^{2}\(ds^{2}_{4}+ds^{2}_{6}\)$, the effective four-dimensional Newton constant is given by $1/G_N^{(4)}=(1/G_N^{(10)})\,L^{6}\int d^{6}x \sqrt g_{6}e^{8\lambda}$, with $16\pi G_{N}^{(10)}=(2\pi)^{7}\ell_{s}^{8}$, following the conventions used in \cite{Guarino:2015jca}. Note in particular that in \cite{Guarino:2015jca} they set $g_s=1$. 
}
\equ{
\frac{1}{G_N^{(4)}}=\frac{6^{5/2}L_{\text{AdS}_{4}}^{6}\vol(Y_{5})}{5\ell_{s}^{8}\pi^{6}}\,.
}

%%%%%%%%%%%%%%%%%
\subsubsection{Entropy}
%%%%%%%%%%%%%%%%%

Putting all this together, we can compute the properly normalized horizon area and obtain the following general formula for the black hole entropy:
\equ{\label{genEntropyIIA}
S_{\rm BH}= (\fg-1)\,\frac{2^{1/3}\,3^{1/6}\,\pi^{3}}{5\,\vol(Y_{5})^{2/3}}\;n^{1/3}\,N^{5/3}\,.
}
Let us specialize this to the two simple cases considered above, namely $\Bbb{CP}^{2}$ and $\Bbb{P}^{1}\times\Bbb{P}^{1}$, with corresponding $Y_{5}$'s equal to $S^{5}$ and $Y^{1,0}$, respectively. We recall that for a $U(N)^{G}$ gauge theory, where each factor has the same Chern-Simons level $k$, the parameter $n$ is related to the Chern-Simons levels by $n=Gk$ \cite{Gaiotto:2009mv}.

For $Y_{5}=S^{5}$ we have $\vol(Y_{5})=\pi^{3}$ and $n=k$ and  \eqref{genEntropyIIA} reads
\equ{\label{entropyS5}
S_{\rm BH}^{S^{5}}=(\fg-1)\, \pi \, 2^{1/3}3^{1/6}5^{-1}N^{5/3}k^{1/3}\,.
}
The dual three-dimensional field theory is a $U(N)$ Chern-Simons theory  with three adjoints superfields $\Phi_i$  and
superpotential $W= \Phi_1[\Phi_2,\Phi_3]$ \cite{Jafferis:2011zi}. The index is explicitly computed in Appendix \ref{app:fieldIIA}; it is given by (\ref{index theorem:attractor2}) and  (\ref{IIAN=4}), where $\Delta_1+\Delta_2+\Delta_3=2\pi$ and $\fn_1+\fn_2+\fn_3=2(1-\fg)$.  The exact R-symmetry corresponds to $\Delta_i=2\pi/3$ and the universal twist to $\fn_i= 2(1-\fg)/3$. Quantization of fluxes requires $\fg -1$ to be an integer multiple of $3$. We then see that the index is given by 
\begin{equation}
\log Z = (\fg -1) \, \pi \, 2^{1/3}3^{1/6}5^{-1}N^{5/3}k^{1/3} ( 1- i/\sqrt{3})\,,
\end{equation}
while the free energy on $S^3$ is given by equation (\ref{indexS32}),
\begin{equation}
F_{S^3} =\pi\, 2^{1/3}3^{1/6}5^{-1}N^{5/3}k^{1/3} ( 1- i/\sqrt{3})\, .
\end{equation}
As expected this leads to $\log Z= (\fg -1) F_{S^3}$, in agreement with our general argument in Section \ref{sec:fieldtheory}. We also note that both $\log Z$ and $F_{S^3}$ are complex but as discussed above Equation \eqref{ident} we should focus on the real part. We then arrive at the following, by now familiar, relation between the black hole entropy and the SCFT partition functions; $S_{\rm BH}={\rm Re} \log Z= (\fg -1) {\rm Re} F_{S^3}$.  

Similarly, for  $Y^{5}=Y^{1,0}$ one has $\vol(Y_{5})=\frac{16\pi^{3}}{27}$, $n=2k$ and \eqref{genEntropyIIA} reads
\equ{\label{entropyY10}
S_{\rm BH}^{Y^{1,0}}=(\fg-1)\, \pi \, 2^{-7/3}3^{13/6}5^{-1}N^{5/3}k^{1/3}\,.
}
The three-dimensional field theory dual is a $U(N)\times U(N)$ gauge theory with four bifundamentals $\Phi_{I}$ with the same quiver and superpotential as the 4d $\mathcal{N}=1$ Klebanov-Witten theory \cite{Klebanov:1998hh}. In this case the exact R-symmetry corresponds to $\Delta_I=\pi/2$ and the universal twist to $\fn_I= (1-\fg)/2$, which requires $\fg$ to be odd. The topologically twisted index, computed explicitly in Appendix \ref{app:fieldIIA}, correctly reproduces the entropy in \eqref{entropyY10}.

\subsubsection{Checking supersymmetry}

As a final consistency check of the supergravity discussion of these new massive IIA black hole solutions here we  show explicitly that the solution in \eqref{flowIIAmetric} and \eqref{eq:uplift2A} preserves two supercharges. The supersymmetry variations of the fermionic fields are given in \cite{Romans:1985tz} and read\footnote{The conventions of this reference are related to ours by $\phi\leftrightarrow-\frac{1}{2}\hat \phi, F_{mnpq}\leftrightarrow \frac{1}{2}\hat F_{mnpq},G_{pqr}\leftrightarrow \hat H_{pqr}, B_{pq} \leftrightarrow \hat B_{pq}$.}
\begin{equation}\label{eq:var}
\begin{split}
\delta \psi_{\mu}=&\,D_{\mu}\epsilon-\frac{1}{32}m \, e^{\frac54 \hat \phi}\Gamma_{\mu}\epsilon -\frac{1}{64} e^{\frac34 \hat \phi}\hat F_{\nu\rho}(\Gamma_{\mu}^{\;\;\nu \rho}-14 \delta_{\mu}^{\;\;\nu}\Gamma^{\rho})\Gamma_{11}\epsilon\\
&+\frac{1}{96}e^{-\frac12 \hat \phi}\hat H_{\nu\rho\lambda}(\Gamma_{\mu}^{\;\;\nu\rho\lambda}-9 \delta_{\mu}^{\;\;\nu}\Gamma^{\rho\lambda})\Gamma_{11}\epsilon+\frac{1}{256}e^{\frac14 \hat \phi}\hat F_{\nu\rho\lambda\tau}(\Gamma_{\mu}^{\;\;\nu\rho\lambda\tau}-\frac{20}{3} \delta_{\mu}^{\;\;\nu}\Gamma^{\rho\lambda\tau})\epsilon\,,\\
\delta \lambda=&\, -\frac{1}{2\sqrt 2}(\partial_{\mu}\hat\phi) \, \Gamma^{\mu}\epsilon-\frac{5}{8\, \sqrt2}\, m\, e^{\frac54 \hat \phi}\epsilon+\frac{3}{16\, \sqrt2}\,  e^{\frac34 \hat \phi}\hat F_{\mu\nu}\Gamma^{\mu\nu}\Gamma_{11}\epsilon\\
&+\frac{1}{24 \, \sqrt 2}\, e^{-\frac12 \hat \phi} \hat H_{\mu\nu\rho}\Gamma^{\mu\nu\rho}\Gamma_{11}\epsilon-\frac{1}{192 \, \sqrt 2}\, e^{\frac14 \hat \phi} \hat F_{\mu\nu\rho\lambda}\Gamma^{\mu\nu\rho\lambda}\epsilon\,.
\end{split}
\end{equation}
Plugging in the background \eqref{flowIIAmetric} and \eqref{eq:uplift2A} and setting these variations to zero we obtain a set of differential equations for $\epsilon$, with the following solution  (see Appendix \ref{sec:AppSUSYIIA} for details): 
\begin{equation}
\epsilon = |g_{tt}|^{1/4} e^{\frac{3}{2}\Gamma^{67}\psi}\mathcal{R}(\theta_1,\theta_2) \eta \, ,
\end{equation}
where $g_{tt}$ the time-time component of the 10d metric in \eqref{flowIIAmetric}. The quantity $\mathcal{R}(\theta_1,\theta_2)$ is an $\alpha$-dependent ``rotation operator,'' defined as
\begin{equation}\label{eq:rotations}
\begin{split}
\mathcal{R}(\theta_1,\theta_2) &\equiv \mathcal{R}_{29}(\theta_1) \mathcal{R}_{1234}(\theta_2) \,; \\
\mathcal{R}_{29}(\theta_1) &\equiv \left(\cos\left(\theta_1/2\right)-\sin\left(\theta_1/2\right)\Gamma^{29}\right) \,, \\
\mathcal{R}_{1234}(\theta_2) &\equiv \left(\cos\left(\theta_2/2\right)-\sin\left(\theta_2/2\right)\Gamma^{1234}\right) \,,
\end{split}
\end{equation}
where the functions $\theta_{1,2}=\theta_{1,2}(\alpha)$ are given in \eqref{eq:angles}, and $\eta$ is a constant ten-dimensional spinor obeying the projection conditions
\equ{\label{projeta}
\Pi_{1}\eta=\Pi_{2}\eta=\Pi_{3}\eta=\Pi_{4}\eta=0\,.
}
Here we have defined the projectors
\eqsn{
\Pi_1 &\equiv \frac{1}{2}\left(1-\Gamma^{5678} \right) \,, && \Pi_2 \equiv \frac{1}{2}\left(1+\Gamma^{3467} \right) \,, \\
\Pi_3 &\equiv \frac{1}{2}\left(1+\Gamma^{134}\right) \,, && 
\Pi_4 \equiv \frac{1}{2}\left(1+\Gamma^{67910}\right) \,.
}
Using \eqref{eq:rotations} one may alternatively  express \eqref{projeta} as  projection conditions on $\epsilon$:
\equ{
\Pi_{1}\epsilon=\Pi_{2}\epsilon=\widehat \Pi_{3}\epsilon=\widehat \Pi_{4}\epsilon=0\,,
}
where we defined the ``rotated projectors,''
\eqs{
\widehat \Pi_3 \equiv \mathcal{R}(\theta_1,\theta_2) \,\Pi_3 \, \mathcal{R}^{-1}(\theta_1,\theta_2) \,, &&
\widehat \Pi_4 \equiv \mathcal{R}(\theta_1,\theta_2) \, \Pi_4 \, \mathcal{R}^{-1}(\theta_1,\theta_2) \,.
}
Similar rotated projectors have appeared in \cite{Pilch:2015dwa,Pilch:2015vha}, albeit in a somewhat different class of supergravity solutions with internal fluxes. Note that the  $\Pi_3$ projector can be understood as a dielectric deformation of the standard D2-brane projector due to the internal fluxes. The four projection conditions in \eqref{projeta} determine that the black hole solution of interest preserves 2 out of the 32 supercharges of the massive IIA supergravity.\footnote{As usual here we discuss only the Poincar\'e type supercharges. For a supersymmetric background with an AdS factor in the metric the number of supercharges is doubled due to the presence of superconformal supercharges.} Note this is half the number of  supercharges of the AdS$_4$ solution of \cite{Guarino:2015jca}, since the $\Pi_2$ projector is absent in that case. 

%%%%%%%%%%%%%%%%%%%%%%%%%%%%%%%%%%%%%
\section{AdS$_2$ horizons in 11d}
\label{AdS2Hor}
%%%%%%%%%%%%%%%%%%%%%%%%%%%%%%%%%%%%%

\global\long\def\vol{\mathrm{vol}}

The discussion thus far has been restricted to the so-called universal twist of 3d $\mathcal{N}=2$ SCFTs and its holographic dual description in terms of a magnetically charged black hole in AdS$_4$. As emphasized, the universal twist is characterized by the fact that the background magnetic flux in the CFT is only along the direction of the 3d superconformal R-symmetry. Most 3d $\mathcal{N}=2$ SCFTs with holographic duals, however, admit continuous flavor symmetries and it is natural to study the behavior of these theories in the presence of magnetic fluxes for these global symmetries, as well as their  holographic duals. In this section we begin exploring this question for a class of 3d $\mathcal{N}=2$ SCFTs arising from M2-branes placed at the tip of a $CY_4$ conical singularity in M-theory. It is well-known that upon backreaction of the branes this setup leads to a supersymmetric Freund-Rubin vacuum of M-theory of the form AdS$_4\times X_7$ where $X_7$ is a Sasaki-Einstein manifold which is the base of the $CY_4$ cone. We have already mentioned a few of these spaces above, for example $M^{1,1,1}$, $Q^{1,1,1}$, and $V^{5,2}$. Many other examples of such Sasaki-Einstein manifolds with explicit metrics are known in the literature (see for example \cite{Gauntlett:2004hh} and references thereof). In addition to the omnipresent Reeb vector, which exists on all such spaces and is dual to the superconformal R-symmetry, these examples of Sasaki-Einstein manifolds typically have additional isometries, which are dual to mesonic flavor symmetries in the dual CFT. In addition, if the seven-dimensional manifold has non-trivial two-cycles (and thus by Poincar\'e duality five-cycles) the dual CFT has continuous baryonic symmetries. After placing the 3d SCFT on $S^1\times \Sigma_\fg$ we can turn on background magnetic fluxes for these mesonic and baryonic flavor symmetries while still preserving supersymmetry.\footnote{From the point of view of the M2-brane picture this corresponds to wrapping the M2-branes on $\Sigma_\fg$ which then by the general result in \cite{Bershadsky:1995qy} automatically implements the topological twist and fibers the $CY_4$ over the Riemann surface.} The holographic dual description of this setup should be given by a magnetically charged supersymmetric black hole which preserves (at least) two supercharges and is asymptotic to the original AdS$_4\times X_7$ background. The near horizon limit of this black hole should be given by a warped product of the form AdS$_2\times_w M_9$ where the manifold $M_9$ is a fibration of the Sasaki-Einstein space $X_7$ over $\Sigma_\fg$. Our goal in this section is to construct explicit examples of such warped AdS$_2$ supersymmetric vacua of eleven-dimensional supergravity.

A few comments on our strategy to attack this technically challenging problem are in order. A classification of warped AdS$_2$ solutions of 11d supergravity was given in \cite{Kim:2006qu,Gauntlett:2007ts}. The result of these papers is that the manifold $M_9$ should be given as a $U(1)$ bundle over an eight-dimensional K\"ahler manifold. The metric on this eight-dimensional space should obey a certain fourth-order non-linear PDE.\footnote{It is curious to note that such K\"ahler manifolds have appeared recently in a seemingly unrelated context in \cite{Ang:2016tvp}.} Finding explicit expressions for such metrics is a challenging problem to tackle  (see however \cite{Gauntlett:2006ns,MacConamhna:2006nb,Donos:2008ug} for some examples) so we employ a different strategy, which closely follows the one in \cite{Benini:2015bwz} where many explicit AdS$_3$ vacua of type IIB supergravity with $(0,2)$ supersymmetry were constructed. To be more concrete we propose an Ansatz for the metric and four-form flux, $G_4$ of the 11d theory and impose the vanishing of the gravitino supersymmetry variation as well as the equations of motion and the Bianchi identity for $G_{4}$. We show that all solutions within our Ansatz are determined by a single function satisfying a nonlinear ODE. This nonlinear equation admits quartic polynomial solutions which lead to a rich class of supergravity solutions. This class contains the near horizon AdS$_2$ region of the universal solution  \eqref{redAnsatz11d} as a special case.

After this brief introduction let us proceed with the construction of our solutions. The eleven-dimensional space is of the form AdS$_{2}\times_{w}M_9$ where $M_9$ is a nine-dimensional manifold given by a seven-dimensional space $M_7$ fibered over a Riemann surface $\Sigma_{\fg}$:
\[
\begin{array}{cccc}
M_7 & \hookrightarrow & M_9&\\ 
 &  & \downarrow & .\\
 &  & \Sigma_{\mathfrak{g}}&
\end{array}
\]
We will take $\Sigma_{\fg}$ to be compact (i.e. with no punctures) and have a constant-curvature metric,\footnote{In principle, this assumption could be relaxed but the analysis is more involved. In addition  the results of  \cite{Anderson:2011cz}  suggest that all the interesting physics in the IR is captured by the constant-curvature metric.}  given by
\equ{
ds_{\Sigma_{\mathfrak{g}}}^{2}=e^{2h(x_{1},x_{2})}\left(dx_{1}^{2}+dx_{2}^{2}\right),
}
with
\equ{\label{defh}
h(x_{1},x_{2})=\begin{cases}
-\log\frac{1+x_{1}^{2}+x_{2}^{2}}{2} & \mathfrak{g}=0\\
\frac{1}{2}\log2\pi & \mathfrak{g}=1\,.\\
-\log x_{2} & \mathfrak{g}>1
\end{cases}
}
The Ansatz for $M_{7}$ will be modelled on a large class of seven-dimensional
Sasaki-Einstein manifolds found and studied in \cite{Gauntlett:2004hh,Martelli:2008rt},
denoted by $Y^{p,q}(\mathcal{B})$, where $\mathcal{B}$, in our case,
is a four-dimensional K\"{a}hler-Einstein manifold that can be either
$\mathbb{CP}^{1}\times\mathbb{CP}^{1}$ or $\mathbb{CP}^{2}$, upon
which the full manifold is constructed. $M_7$ is then spanned by the four coordinates on $\mathcal{B}$
and $(y,\beta,\psi)$, the latter two being angles. In particular,
$\psi$ is the angle associated to the Reeb vector of $M_{7}$. Thus, we employ an eleven-dimensional metric Ansatz which has explicit AdS$_2$, $\Sigma_\fg$ and $\mathcal{B}$ factors as well as two $U(1)$ isometry directions, $\beta$ and $\psi$, which are fibered over $\Sigma_\fg$ and $\mathcal{B}$. With these assumptions we arrive at the following general metric Ansatz:
\begin{equation}
ds_{11}^{2}=f_{1}^{2}\,ds_{\text{AdS}_{2}}^{2}+f_{2}^{2}\,ds_{\Sigma_{\mathfrak{g}}}^{2}+f_{3}^{2}\,dy^{2}+f_{4}^{2}\,(D\beta)^{2}+f_{5}^{2}\,ds_{{\cal B}}^{2}+f_{7}^{2}\,(D\psi)^{2},\label{AdS2Hor_metricAnsatz}
\end{equation}
where $f_{n}=f_{n}(y)$ are functions of the coordinate $y$ only, and 
\begin{equation}
\label{defDpsibeta}
D\beta\equiv d\beta+cA+bA_{{\cal B}}\,, \qquad D\psi\equiv d\psi+aA_{{\cal B}}+\tilde{c}A-f_{8}(y)D\beta\,,
\end{equation}
with $A=A_{1}(x_{1},x_{2})\,dx_{1}+A_{2}(x_{1},x_{2})\,dx_{2}$ a connection on the Riemann surface and $A_{\mathcal{B}}$ a connection on $\cB$, satisfying,
\begin{equation}
dA=\vol_{\Sigma_{\mathfrak{g}}}\,, \qquad dA_{\mathcal{B}}=J_{\mathcal{B}}\,,
\label{AdS2Hor_ConnectionRels}
\end{equation}
with $J_{\mathcal{B}}$ the K\"ahler form on $\cB$. The most general Ansatz for the 4-form compatible with the symmetries of the problem reads
\begin{alignat}{1}
G_{4}  =&\, f_1^2 \vol_{\text{AdS}_{2}}\wedge\left(f_2^2 g_{1}\vol_{\Sigma_{\mathfrak{g}}}+f_3 f_4 g_{2}dy\wedge D\beta+f_4 f_7 g_{3}D\beta\wedge D\psi+ f_3 f_7 g_{4}dy\wedge D\psi+f_5^2 g_{5}J_{\mathcal{B}}\right)\nonumber \\
 & +f_2^2 \vol_{\Sigma_{\mathfrak{g}}}\wedge\left( f_3 f_4 i_{1}dy\wedge D\beta + f_5^2 i_{2}J_{\mathcal{B}} + f_4 f_7 i_{4}D\beta\wedge D\psi + f_3 f_7 i_{5}dy\wedge D\psi\right)\label{AdS2Hor_4FormAnsatz}\\
 & +f_5^2 J_{\mathcal{B}}\wedge\left(f_3 f_4 l_{1}dy\wedge D\beta+f_5^2 l_{2} J_{\mathcal{B}} + f_4 f_7 l_{3}D\beta\wedge D\psi+ f_3 f_7 l_{4}dy\wedge D\psi\right)\,,\nonumber 
\end{alignat}
where $g_{n}=g_{n}(y),i_{n}=i_{n}(y)$, and $l_{n}=l_{n}(y)$ are  functions of the coordinate $y$ only. 

Before we describe the solutions, let us discuss the interpretation of various terms in the Ansatz. Since the Reeb vector $\partial_{\psi}$ corresponds to the $U(1)_{R}$ R-symmetry of the 3d $\cN=2$ field theory dual, the $\tilde c A$ term in $D\psi$ thus corresponds to the part of the topological twist along the three-dimensional superconformal R-symmetry. The isometry $\partial_{\beta}$, on the other hand, is dual to a flavor symmetry and thus the  $cA$ term in $D\beta$ corresponds to flavor flux in the dual field theory (which should be set to zero in the special case of the universal twist). The parameters $a,b$ specify the structure of the fibration over $\cB$. From the black hole perspective, since the reduction along $D\psi$ and $D\beta$ leads to a graviphoton and a vector field, respectively, the coefficients $c,\tilde c$ are related to the magnetic charges of the black hole under these two gauge fields. Other charges may arise in four dimensions from the reduction of the 4-form, i.e., Betti multiplets. These arise from the terms proportional to elements of $H^{2}(M_7)$
and $\vol_{\Sigma_{\mathfrak{g}}}$ and correspond to a background for a baryonic symmetry in the dual picture; we discuss a class of examples of this in section~\ref{AdS2Hor-BaryonicFluxes}. It is worth pointing out that the map between the parameters in our supergravity Ansatz and quantities in the dual CFT could in general be complicated by a possible mixing between the three-dimensional superconformal R-symmetry and flavor and baryonic symmetries as one flows to the IR. Thus, one should take the field theory interpretation of the supergravity parameters with a healthy dose of skepticism.

Plugging the Ansatz \eqref{AdS2Hor_metricAnsatz} and \eqref{AdS2Hor_4FormAnsatz} into the BPS equations and the equations of motion and Bianchi identity for the four-form: $dG_{4}=d\ast_{11} G_{4}+ \frac{1}{2}G_4\wedge G_4=0$, one obtains a set of differential equations, analyzed in detail in  Appendix~\ref{Appendix11D}. As discussed there it is convenient to introduce a new basis of functions $F_{i}(y)$, related to the original functions $f_{i}(y)$  by \eqref{AdS2Hor_FBasis}. Here we give the final result of our analysis, which gives a metric
\begin{alignat}{1}
ds_{11}^{2}  =&\,\(\frac{F_{2}^{2}F_{3}^{4}}{F_{1}^{2}}\)^{1/3}\,ds_{\text{AdS}_{2}}^{2}+\(\frac{F_{1}F_{2}^{2}}{F_{3}^{2}}\)^{1/3}\,ds_{\Sigma_{\mathfrak{g}}}^{2}+K\frac{\(F_{1}F_{2}^{2}F_{3}^{4}\)^{1/3}}{F_{5}}\,dy^{2}\nonumber \\
 & +KF_{5}\(\frac{F_{1}}{F_{2}^{4}F_{3}^{8}}\)^{1/3}\,(D\beta)^{2}+\(\frac{F_{1}F_{3}}{F_{2}}\)^{1/3}\,ds_{{\cal B}}^{2}\label{AdS2Hor_AltAnsatzMain}\\
 & +4\omega_{1}^{2}\(\frac{F_{2}^{2}F_{3}^{4}}{F_{1}^{2}}\)^{1/3}\left(d\psi+aA_{{\cal B}}+\tilde{c}A-\frac{F_{6}}{F_{2}F_{3}^{2}}\,D\beta\right)^{2}\,\nonumber 
\end{alignat}
and 4-form
\begin{alignat}{1}
G_{4}  =&\,\vol_{\text{AdS}_{2}}\wedge\left[\frac{2F_{1}F_{2}-2\kappa F_{2}F_{3}^{2}+cF_{5}'}{2F_{1}}\,\vol_{\Sigma_{\mathfrak{g}}}+\left(2\omega_{1}\frac{F_{2}F_{3}^{2}}{F_{1}}\left(\frac{F_{6}}{F_{2}F_{3}^{2}}\right)'+K\right)\,dy\wedge D\beta\right.\nonumber\\
 & \left.-2\omega_{1}\frac{\left(cKF_{3}^{2}+2bKF_{2}F_{3}\right)F_{1}-F_{2}F_{3}^{2}F_{1}'}{F_{1}^{2}}\,dy\wedge D\psi+\frac{2F_{1}F_{3}-4qF_{2}F_{3}^{2}+bF_{5}'}{2F_{1}}\, J_{\mathcal{B}}\right]\label{AdS2Hor_G4Solution}\\
 & +\frac{F_{2}\mathcal{I}_{0}}{F_{3}^{2}}\left[\vol_{\Sigma_{\mathfrak{g}}}\wedge\left(-2dy\wedge D\beta+J_{\mathcal{B}}\right)+J_{\mathcal{B}}\wedge\left(dy\wedge D\beta-J_{\mathcal{B}}\right)\right]\,,\nonumber
\end{alignat}
where $\omega_{1}$ and $\cI_{0}$ are integration constants, the functions $F_{2},F_{3}$ are linear in $y$ and $F_{1},F_{6}$ are given in terms of $F_{2},F_{3}$ and the function $F_{5}$ by:
\begin{equation}
\begin{split}F_{1} & =-\frac{F_{5}''}{2K}+\kappa F_{3}^{2}+4qF_{2}F_{3}\,,\\
F_{2} & =cKy+S_{2}\,,\\
F_{3} & =bKy+S_{3}\,,\\
F_{6} & =\frac{F_{5}'-4\omega_{2}F_{2}F_{3}^{2}}{4\omega_{1}}\,,
\end{split}
\label{AdS2Hor_FSummary}
\end{equation}
with $\{S_{2},S_{3},\omega_{2},K\}$  integration constants\footnote{In Appendix~\ref{Appendix11D-Singularities} we show that actually $\omega_{2}$ can be always
set to zero without loss of generality. Furthermore, $\omega_1$ and $K$ can be set to convenient non-zero values.} and $q$ is related to the Ricci curvature of $\mathcal{B}$ by:
\equ{
q=\frac{R_{\mathcal{B}}}{8}\,.
}
The entire class of solutions is thus characterized by the single function $F_{5}$ which, as shown in Appendix~\ref{Appendix11D}, satisfies the second-order nonlinear ODE given in $\eqref{AdS2Hor_F5Eq}$ coming from the Bianchi identity  for $G_{4}$, which for $b\neq 0$ reads
\eqss{
\frac{1}{KF_{2}F_{3}^{2}}\left[-F_{5}'^{2}+2F_{5}\left(F_{5}''-2K\left(4qF_{2}F_{3}+\kappa F_{3}^{2}\right)\right)\right]-\frac{4\mathcal{I}_{0}^{2}\left(3bcKy+bS_{2}+2cS_{3}\right)}{b^{3}KF_{3}^{2}}\\
+\frac{16}{3}Kqy^{2}\left(b\kappa Ky+cKqy+3\kappa S_{3}+3qS_{2}\right)+p_{0}+p_{1}y=0\,,\label{AdS2Hor_F5EqMain}
}
with $p_{0},p_{1}$  two more integration constants. 
Finally, there are some constraints on $a,b,c,\tilde{c}$:
\equ{\label{eq:caconstr}
c\, \omega_{2}+\tilde{c}\, \omega_{1}  =\frac{\kappa}{2}\,,\qquad 
b\, \omega_{2}+a\, \omega_{1}  =q\,.
}
Thus, the most general solution within our Ansatz is characterized by solutions to the nonlinear second-order ODE \eqref{AdS2Hor_F5EqMain}.\footnote{For $b=0$ the ODE follows from  the general Equation \eqref{AdS2Hor_F5Eq}.} Although we have not found the most general solution,  one can show that the most general {\it polynomial} solution is at most quartic, i.e.,  
\equ{\label{eq:F5pol}
F_{5}(y)=\sum_{n=0}^{4}\alpha_{n}\, y^{n}\,.
}
Plugging this into \eqref{AdS2Hor_F5EqMain} leads to a set of algebraic equations for the coefficients $\alpha_{n}$ and the other integration constants specifying the solution, given in \eqref{constalphas}.
Solutions to this set of algebraic constraints leads to a rich family of local AdS$_{2}$ solutions. As we discuss next, it contains the universal solution as a special case, as well as interesting generalizations of it by the addition of mesonic and baryonic flavor fluxes.

%%%%%%%%%%%%%%
\subsection{Universal twist\label{AdS2Hor-UniversalTwist}}
%%%%%%%%%%%%%%

The universal solution is characterized by the absence of flavor fluxes.
To this end we will set $c=\mathcal{I}_0=0$.
We will also assume $b\neq 0$, leaving the case $b=0$ to the next section, as this case admits a more general solution.
One can show that there are no solutions for $\kappa=0$, but the values $\kappa=\pm1$ are allowed by the constraints.
We will now discuss the solution for $\kappa=-1$ and discuss the case $\kappa=1$ in Section \ref{sec:AdS2xS3xT6}.
Plugging in the polynomial Ansatz \eqref{eq:F5pol} for $F_5$ into the Bianchi constraint \eqref{AdS2Hor_F5EqMain} fixes $\{p_0,p_1,\alpha_1,\alpha_2,\alpha_3,\alpha_4\}$, leaving $\alpha_0$ undetermined.
There are in fact two solutions to the Bianchi constraint, but only one obeys the metric positivity constraint $F_1>0$.
The parameters $a,\tilde{c}$ are fixed by \eqref{eq:caconstr} to be
\[
a=\frac{q}{\omega_1}\,,\qquad\tilde{c}=\frac{\kappa}{2\omega_1}\,.
\]
The dependence of the solution on the parameters $\omega_1,\omega_2,K$ and $S_3$ can be removed by rescaling and shifting $y,\beta$ and $\psi$.
It is convenient to define $S_2 \equiv L^3/32$ and to perform a rescaling $\beta'=2\beta,\psi'=(\omega_1/2)\psi $
so that now $D\beta'=d\beta'+2cA+2bA_{{\cal B}}$, and define the polynomial
\[
Q_{2}(x)\equiv3x^{2}+2x+1\,.
\]
The metric \eqref{AdS2Hor_AltAnsatzMain} can then be written as  
\eqs{\nonumber
ds_{11}^{2} =&\,L^{2}\left[\frac{1}{4}\left(\frac{1}{4}ds_{\text{AdS}_2}^{2}+\frac{1}{2} ds^2_{\Sigma_g}\right)+\frac{3}{4}\frac{\left(1-c_{Y^{p,q}}y\right)^{2}}{a_{Y^{p,q}}-y^{2}Q_{2}(1-c_{Y^{p,q}}y)}\,dy^{2}\right.\\
 & +\frac{q^{2}}{48}\frac{a_{Y^{p,q}}-y^2\, Q_{2}(1-c_{Y^{p,q}}y)}{\left(1-c_{Y^{p,q}}y\right)^{2}}(D\beta')^{2}+\frac{q}{4}\left(1-c_{Y^{p,q}}y\right)ds_{{\cal B}}^{2}\\ \nonumber
 & \left.+\left(d\psi'+\frac{q}{2}A_{{\cal B}}-\frac{1}{4}A+\frac{q}{4}yD\beta'\right)^{2}\right],
}
where, inspired by the notation in \cite{Gauntlett:2004hh}, we defined $a_{Y^{p,q}}$ to be a constant related to $\alpha_{0}$ and $c_{Y^{p,q}}=-b$. This metric has exactly the form of the near horizon limit of the universal metric in \eqref{redAnsatz11d}.
The seven-dimensional manifold parametrized by $x,\mathcal{B},\psi',\beta'$ at a fixed point on the Riemann surface, coincides with the seven-manifold described in \cite{Gauntlett:2004hh}.\footnote{In their notation we have $\Lambda=8, \lambda=2q$.} The corresponding 4-form reads
\begin{equation}
G_4 = \left(\frac{L}{4}\right)^3 \vol_{\text{AdS}_{2}}\wedge \left(3 \text{vol}_{\Sigma_g} + q \, dx \wedge D\beta'+2 q\, (1-c_{Y^{p,q}}) J_\mathcal{B} \right)\, .
\end{equation}
Deformations of this solution with $c\neq0$, which we do not analyze in full detail here, should thus describe the twisted compactification of the corresponding SCFTs with flavor flux. 

%%%%%%%%%%%%%%%%%%%%%%%%%%%%%%%%%
\subsection{AdS$_2 \times S^3 \times T^6$}\label{sec:AdS2xS3xT6}
%%%%%%%%%%%%%%%%%%%%%%%%%%%%%%%%%

As mentioned in Section \ref{AdS2Hor-UniversalTwist}, there is another solution within our Ansatz with $b\neq 0$, $\kappa=1$ and $c=0$, which reads
\begin{equation}
ds^2 = L^2 \left( ds^2_{\text{AdS}_2}+ ds^2_{S^2} + \frac{3}{4} \frac{x^2\,dx^2}{C+x^3}+\frac{4}{3} \frac{C+x^3}{x^2}\, q^2\, (d\beta'+A_{\mathcal{B}})^2 + x \, q \, ds^2_{\mathcal{B}}+(d\psi'+A)^2 \right) \, ,
\end{equation}
where $x$ is related to $y$ by a shift and rescaling, $\beta'= \beta/b$, $\psi'= 2 (\omega_1 \psi-q \beta')$, and $C$ is a constant. The flux for this solution is given by
\begin{equation}
G_4 = L^3 \, \text{vol}_{\text{AdS}_2} \wedge \left( dx \wedge d\beta'+ x q J_\mathcal{B}\right) \, .
\end{equation}
For $\mathcal{B}=\mathbb{CP}^2$ and $C=0$, the solution becomes the well-known AdS$_2\times S^3\times T^6$ background of 11d supergravity. To see this note that the $S^{2}$ combines with $(d\psi'+A)^2$ into an $S^3$, and $(D\beta')^2$ combines with $ds^2_{\mathbb{CP}^2}$ into an $S^5$.
Then, with the coordinate transformation $x=\rho^2$, we recognize a six-dimensional metric which is simply $\mathbb{R}^6$, written as a cone over $S^5$. We can then periodically identify the coordinates on this $\mathbb{R}^6$ to obtain $T^6$.\footnote{For $\cB=\Bbb P^{1}\times \Bbb P^{1}$ and $C=0$ one obtains a cone over the conifold $Y^{1,0}$. The resulting six-dimensional manifold cannot be made compact and regular.} For $C\neq 0$, instead, the internal manifold is singular and  we thus discard the solution.

%%%%%%%%%%%%%%%%
\subsection{Baryonic fluxes\label{AdS2Hor-BaryonicFluxes}}
%%%%%%%%%%%%%%%%

As mentioned above, in the special case $b=0$ there is a more general solution  to the constraints imposed by Bianchi, even while keeping $c=0$ (i.e., no mesonic fluxes). Note that setting $b=0$ and using the results in \cite{Gauntlett:2004hh} amounts to limiting the discussion to the Sasaki-Einstein manifolds $Q^{1,1,1}$ and $M^{1,1,1}$, and orbifolds thereof. As discussed in detail below, these solutions are characterized by the presence of 4-form flux through non-trivial cycles in $M_9$. This, in four dimensions, means that the corresponding black hole is charged under the Betti multiplets and thus the dual field theory is twisted by baryonic flux, in addition to R-symmetry flux. These charges may be of electric or magnetic type, depending on whether the flux is through AdS$_2$ or $\Sigma_{\mathfrak{g}}$. Here we analyze the general case in which they are both present, leaving to the following paragraphs a more detailed presentation of the purely electric and purely magnetic solutions.\footnote{Black holes charged under  Betti multiplets have been numerically found in  \cite{Halmagyi:2013sla} in four-dimensional supergravity in the particular case of the homogeneous manifolds $Q^{1,1,1}$ and $M^{1,1,1}$  using the consistent truncation in \cite{Cassani:2012pj}.}  

A general result that follows when we set $b=c=0$, is that the functions $f_1$ and $f_2$ are constant.\footnote{This follows from the constraint \eqref{AdS2Hor_F5Eq}, which implies that $F_{5}$ is at most a polynomial of
degree two, \eqref{AdS2Hor_FSummary} and the definitions \eqref{AdS2Hor_FBasis}. } Let us denote these by
\equ{\label{AdS2Hor_BaryonicProof1}
f_{1}\equiv L\,,\qquad f_{2} \equiv \sqrt u \, L\,,
}
with $u,L>0$. Using \eqref{AdS2Hor_FSummary} and  the definitions \eqref{AdS2Hor_FBasis} gives the constraint on $\alpha_2$:
\begin{equation}
F_{5}''=2K\left(\left(\kappa-u\right)S_{3}^{2}+4qS_{2}S_{3}\right)=2\alpha_{2}\,.\label{AdS2Hor_BaryonicProof2}
\end{equation}
On the other hand, the parameters $\alpha_0$ and $\alpha_1$ are not constrained, but we will assume that $\alpha_0 \alpha_2<0$.
Then, the part of the metric involving $y$ and $\beta$, namely
\[
K\frac{uL^{2}S_{3}^{2}}{F_{5}}dy^{2}+KF_{5}\frac{1}{uL^{4}S_{3}^{2}}d\beta^{2}\;,
\]
can  be recast, via rescaling and shifts of the coordinates $y$ and $\beta$, into the metric of a two-sphere with a certain radius $L\sqrt{v}$:
\[
vL^2\left(\frac{1}{1-y^2}dy^{2}+(1-y^2)d\beta^{2}\right).
\]
This choice of coordinates is equivalent to setting
\begin{equation}
K=vL^{3},\quad\alpha_{2}=-S_{2}S_{3}^{2}\,,\quad\alpha_{0}=-\alpha_{2}\,,\quad\alpha_{1}=0\,.
\label{AdS2Hor_BaryonicProof4}
\end{equation}
Combining \eqref{AdS2Hor_BaryonicProof4} with \eqref{AdS2Hor_BaryonicProof2}, we can determine $S_{3}$
in terms of the other parameters:
\begin{equation}
S_{3}=-\frac{4qS_{2}K}{S_{2}+K\left(\kappa-u\right)}\,.\label{AdS2Hor_BaryonicProof3}
\end{equation}
It is convenient to write $\mathcal{I}_0=wS_{3}^{2}$, with $w$ a real parameter. Then, the general solution reads
\begin{eqnarray}
\label{AdS2Hor_BaryonicGenSol} \nonumber
&&ds_{11}^{2}  =\, L^{2}\left(ds_{\text{AdS}_{2}}^{2}+u\,ds_{\Sigma_{\mathfrak{g}}}^{2}+v\,ds_{S^{2}}^{2}+\frac{4quv}{-u+v(u-\kappa)}ds_{{\cal B}}^{2}+\left(d\psi+2q\, A_{{\cal B}}+\kappa \, A+y\, d\beta\right)^{2}\right)\,,\\ 
&&G_{4}  =\,L^{3}\,\vol_{\text{AdS}_{2}}\wedge\left(\left(u-\kappa\right)\vol_{\Sigma_{\mathfrak{g}}}+\left(v-1\right)dy\wedge d\beta+2q\frac{u+v(u+\kappa)}{-u+v\left(u-\kappa\right)}\,J_{{\cal B}}\right)\\ \nonumber
 &&\qquad  \quad +\, uwL^{3}\left[\vol_{\Sigma_{\mathfrak{g}}}\wedge\left(-2dy\wedge D\beta+J_{\mathcal{B}}\right)+J_{\mathcal{B}}\wedge\left(dy\wedge D\beta-J_{\mathcal{B}}\right)\right]\,.
\end{eqnarray}
The three parameters  $u,v,w$ are not independent, but are constrained by the Bianchi identity to satisfy: 
\begin{equation}
\frac{4q^{2}\left(-3\kappa^{2}v^{2}+2\kappa u(v-1)v+u^{2}(v-1)(v+3)\right)+3w^{2}(\kappa v-uv+u)^{2}}{(\kappa v-uv+u)^{4}}=0\,.
\label{AdS2Hor_UVWConstraint}
\end{equation}
We thus obtain a two-parameter family of deformations of the near-horizon geometry of the universal black hole associated to either the $Q^{1,1,1}$ or $M^{1,1,1}$ 3d SCFT, according to the choice of $\cB$. The black hole entropy reads:
\equ{\label{entropy baryonic}
S_{\text{BH}}=\frac{u}{8\pi}\,\vol_{\Sigma_{\fg}} \, F_{S^{3}}\,.
}
%

%%%%%%%%%%%%%
\subsubsection{Electric baryonic charges}
%%%%%%%%%%%%%

In this section, we consider a purely electric baryonic flux and therefore  set $w=0$, to eliminate any magnetic baryonic fluxes that could arise from the terms proportional to $\vol_{\Sigma_{\mathfrak{g}}}$ in the third line  of \eqref{AdS2Hor_BaryonicGenSol}. Setting $w=0$ in the constraint  \eqref{AdS2Hor_UVWConstraint} we find that the resulting metric is acceptable only for $\kappa=-1$ and the solution reads:
\begin{equation}
\begin{split}ds_{11}^{2} & =L^{2}\left(ds_{\text{AdS}_{2}}^{2}+u\,ds_{\Sigma_{\mathfrak{g}}}^{2}+v\,ds_{S^{2}}^{2}+\frac{4quv}{v-u+uv}\, ds_{{\cal B}}^{2}+\left(d\psi+2qA_{{\cal B}}- A+y\, d\beta\right)^{2}\right)\,,\\
G_{4} & =L^{3}\,\vol_{\text{AdS}_{2}}\wedge\left(\left(u+1\right)\vol_{\Sigma_{\mathfrak{g}}}+\left(v-1\right)dy\wedge d\beta+2q\,\frac{u-v+uv}{v-u+uv}\,J_{{\cal B}}\right)\,.
\end{split}
\label{AdS2Hor_BaryonicEFluxSol}
\end{equation}
The corresponding entropy, given by \eqref{entropy baryonic}, reads
\equ{
S_{\text{BH}}=\frac{u}{2}\,(\fg-1)\, F_{S^{3}}\,.
} 
We note that in the special case $u=v=2$ the solution reduces to the universal twist of  $M^{1,1,1}$ or $Q^{1,1,1}$. The latter was first derived in \cite{Donos:2012sy} (see Equation (3.25) there). To better understand the physical interpretation of the deformation consider the reduction of this eleven-dimensional background to four dimensions. The terms in the 4-form proportional
to $H^{2}(M_7)$ give rise to four-dimensional Betti vector multiplets.
Since $G_{4}$  in \eqref{AdS2Hor_BaryonicEFluxSol} is proportional to $\vol_{\text{AdS}_{2}}$, the black hole
is {\it electrically} charged under this vector field, as measured by the flux of $\ast_{11}\,G_{4}$ through
non-trivial 7-cycles.

Let us take for example $Q^{1,1,1}$. This space is a fibration of $\psi$ over the product of three two-spheres, which we denote by  $\mathcal{S}_{1},\mathcal{S}_{2},\mathcal{S}_{3}$. The generators of $H^{2}(Q^{1,1,1})$
are then given by
\begin{equation}
h_{1}=\vol_{\mathcal{S}_{1}}-\vol_{\mathcal{S}_{2}}\,,\qquad h_{2}=\vol_{\mathcal{S}_{2}}-\vol_{\mathcal{S}_{3}}\,,
\label{AdS2Hor_Q111H2}
\end{equation}
since\footnote{Here by $D\psi$ we mean $d\psi+A_{\text{KE}_{6}}$. }
\[
\vol_{\mathcal{S}_{1}}+\vol_{\mathcal{S}_{2}}+\vol_{\mathcal{S}_{3}}=d(D\psi)
\]
is trivial. Define now the 7-cycles $C_{1},C_{2},C_{3}$ as
the submanifolds obtained by fixing a point in AdS$_{2}$ and $\mathcal{S}_{1},\mathcal{S}_{2},\mathcal{S}_{3}$, respectively. Cohomology suggests that the
non-trivial 7-cycles on which we should integrate $\ast_{11}\,G_{4}$ are
\[
H_{1}=C_{1}-C_{2}\,,\qquad H_{2}=C_{2}-C_{3}\,.
\]
Denoting the sphere spanned by $(y,\beta)$  by $\mathcal{S}_{1}$, it is easy to see that only the Betti multiplet associated to $H_{1}$ has a non-vanishing
flux,\footnote{\label{footbary}Since $b_{2}(Q^{1,1,1})=2$ one should expect a more general solution with two electric baryonic fluxes. The fact that we only see one flux here can be understood as follows. Recall that in our Ansatz \eqref{AdS2Hor_metricAnsatz},  \eqref{AdS2Hor_4FormAnsatz}  we have assumed the base $\cB$ is a K\"ahler-Einstein manifold. This implies that $\vol_{\mathcal{S}_{2}}$ and $\vol_{\mathcal{S}_{3}}$ cannot appear in an arbitrary combination, but instead are always added so as to give $J_{\mathcal{B}}$. If one relaxes the Einstein condition on $\cB$, the resulting equations are almost identical to the ones presented here and their solutions allow for both baryonic fluxes to be present at once. In contrast, since $b_{2}(M^{1,1,1})=1$ our Ansatz captures the most general baryonic solution in that case.} due to the presence of $J_{\mathcal{B}}=\vol_{\mathcal{S}_{2}}+\vol_{\mathcal{S}_{3}}$
in $G_{4}$:
\begin{alignat}{1}
\mathcal{F}_{1} & \equiv\int_{H_{1}}\ast_{11}\,G_{4}=-64\, \pi^{3}\, (\mathfrak{g}-1)\, \Delta_{\psi}\, L^{6}\, \frac{u v (u (v-1)-v) (u (v-3)+v)}{(u (v-1)+v)^2}\,\label{AdS2Hor_ElFlux},\\
\mathcal{F}_{2} & \equiv\int_{H_{2}}\ast_{11}\, G_{4}=0\,\nonumber,
\end{alignat}
where $\Delta_{\psi}$ is the period of $\psi$. Since Betti multiplets are dual to baryonic symmetries in the field theory, for general values of the parameter $v>0$ (or $u>0$), the solution describes the twisted compactification with baryonic fluxes. In the special case $u=v=2$ one has $\mathcal{F}_{1}=0$ and there is no baryonic flux. Indeed, the 4-form  can be written as
\[
G_{4}=L^{3}\,\vol_{\text{AdS}_{2}}\wedge\left(4\, \vol_{\Sigma_{\mathfrak{g}}}+d\,(D\psi)\right)\,,
\]
and only a trivial cocycle of $M_7$ appears.

The solution described here is reminiscent of the AdS$_3$ solutions of type IIB supergravity presented in Section~2.2 of \cite{Benini:2015bwz} (see also \cite{Amariti:2016mnz} for a five-dimensional perspective on this solution), describing  the twisted compactification of the Klebanov-Witten theory (and its $\Bbb Z_{p}$ orbifolds) with magnetic baryonic flux. 

%%%%%%%%%%
\subsubsection{Magnetic baryonic charges}
\label{subsubsec:mag}
%%%%%%%%%%

Magnetic baryonic charges can be obtained by a non-zero $w$. For concreteness, we present the case $\mathcal{B}=\mathbb{CP}^{1}\times\mathbb{CP}^{1}$ (the $\mathbb{CP}^{2}$ case is analogous) and we analyze the solution with a purely magnetic baryonic flux  and thus impose that $\eqref{AdS2Hor_ElFlux}$ vanishes. Combining this constraint with  \eqref{AdS2Hor_UVWConstraint} determines $u$ and $v$ in terms of $w$. In contrast to the purely electric case, there are solutions for all $\kappa=\{0,\pm1\}$.

For $\kappa=-1£ì$ we find 
\begin{equation}
\begin{split}ds_{11}^{2} & =\, L^{2}\left(ds_{\text{AdS}_{2}}^{2}+\left(3w^2+2\right)\, ds_{\Sigma_{\mathfrak{g}}}^{2}+\frac{3w^2+2}{3w^2+1}\, ds_{S^{2}}^{2}+\left(3w^2+2\right)ds_{{\cal B}}^{2}+D\psi^{2}\right)\,,\\
G_{4} & =\,L^{3}\,\vol_{\text{AdS}_{2}}\wedge\left(\left(3w^2+3\right)\vol_{\Sigma_{\mathfrak{g}}}+\frac{1}{3w^2+1}\,dy\wedge D\beta+\left(3w^2+1\right)J_{\mathcal{B}}\right)\\
& + L^{3}\left(3w^2+2\right)w\left[\vol_{\Sigma_{\mathfrak{g}}}\wedge\left(-2dy\wedge d\beta+J_{\mathcal{B}}\right)+J_{\mathcal{B}}\wedge\left(dy\wedge d\beta-J_{\mathcal{B}}\right)\right]\,.
\end{split}
\label{AdS2Hor_BaryonicMFluxSol}
\end{equation}
The universal solution is recovered for $w=0$. For any other value the baryonic charges are given by the integral of $G_4$ through the non-trivial 4-cycles of $M_9$, whose form is suggested by $\eqref{AdS2Hor_Q111H2}$ and schematically reads
\[
\tilde{H}_1 \equiv \Sigma_{\mathfrak{g}}\times\mathcal{S}_1-\Sigma_{\mathfrak{g}}\times\mathcal{S}_2\,,\qquad \tilde{H}_2 \equiv \Sigma_{\mathfrak{g}}\times\mathcal{S}_2-\Sigma_{\mathfrak{g}}\times\mathcal{S}_3\,.
\]
The result for the fluxes is
\begin{alignat}{1}
\mathcal{\tilde{F}}_{1} & \equiv\int_{\tilde{H}_{1}}G_{4}=-48\,|\mathfrak{g}-1|\,\pi^{2}\,L^{3}\,\left(3w^2+2\right)w\,\label{AdS2Hor_MagFlux},\\
\mathcal{\tilde{F}}_{2} & \equiv\int_{\tilde{H}_{2}}G_{4}=0\,.\nonumber
\end{alignat}
The entropy reads
\equ{
S_{\text{BH}}= \frac12 \,(3 \,w^2+2) \, (\fg-1)\,   F_{S^{3}}\,.
}

For $\kappa=1$ and $|w|>1$, we find
\begin{equation}
\begin{split}ds_{11}^{2} & =L^{2}\left(ds_{\text{AdS}_{2}}^{2}+\frac{w^{2}+2}{w^{2}-1}\,ds_{S^2}^{2}+(w^{2}+2)\,ds_{S^{2}}^{2}+(w^{2}+2)\,ds_{{\cal B}}^{2}+D\psi^{2}\right)\,,\\
G_{4} & =L^{3}\, \vol_{\text{AdS}_{2}}\wedge\left(\frac{3}{w^{2}-1}\,\vol_{S^2}+\frac{3}{w^{2}-1}\,d\theta_{1}\wedge d\phi_{1}+(w^{2}+1)\,J_{{\cal B}}\right)\\
& +w\,\frac{w^{2}+2}{w^{2}-1}\,L^{3}\,\left[\,\vol_{S^2}\wedge\left(-2\,dy\wedge d\beta+J_{\mathcal{B}}\right)+J_{\mathcal{B}}\wedge\left(dy\wedge d\beta-J_{\mathcal{B}}\right)\right]\,.
\end{split}
\label{AdS2Hor_BaryonicMFluxSolAlt}
\end{equation}
The resulting magnetic flux along $\tilde{H}_1$ is
\begin{equation}
\mathcal{\tilde{F}}_{1} =-48\pi^{2}L^{3}\, w\, \frac{w^{2}+2}{w^{2}-1}\,,
\end{equation}
and the entropy is given by
\begin{equation}
S_{\text{BH}}=\frac{1}{2}\,\frac{w^2+2}{w^2-1}\,F_{S^{3}}\,.
\end{equation}
For the special value $w=\sqrt{2}$ we obtain a solution analogous to the one presented in \cite{Donos:2012sy} (see Equation (3.16) there), where the other Betti multiplet was turned on. As discussed in Footnote \ref{footbary}, the absence of the second Betti multiplet is a consequence of our Ansatz. 

Finally, for $\kappa=0$ the solution can be expressed in terms of $u$ alone and reads\footnote{The constraint \eqref{AdS2Hor_UVWConstraint} has two acceptable solutions leading to the two signs in the last line of \eqref{AdS2Hor_MagBarTorusSol}. This choice of sign amounts to the sign of the magnetic baryonic charge.}
\begin{equation}
\label{AdS2Hor_MagBarTorusSol}
\begin{split}ds_{11}^{2} & =L^{2}\left(ds_{\text{AdS}_{2}}^{2}+u\,ds_{T^{2}}^{2}+3\,ds_{S^{2}}^{2}+3\,ds_{{\cal B}}^{2}+\left(d\psi+A_{{\cal B}}+y\,d\beta\right)^{2}\right)\,,\\
G_{4} & =L^{3}\,\vol_{\text{AdS}_{2}}\wedge\left(u\,\vol_{T^{2}}+2\,dy\wedge d\beta+2J_{{\cal B}}\right)\\
 &\quad \pm u\,L^{3}\,\left[\,\vol_{T^{2}}\wedge\left(-2\,dy\wedge D\beta+J_{\mathcal{B}}\right)+J_{\mathcal{B}}\wedge\left(dy\wedge D\beta-J_{\mathcal{B}}\right)\,\right]\,,
\end{split}
\end{equation}
By fixing $u$ one recovers a solution analogous to the one found in \cite{Donos:2012sy} (see Equation (2.17) there). The entropy and baryonic flux are given by 
\begin{equation}
S_{\text{BH}}=\frac{1}{4}\,u\, F_{S^{3}}\,, \qquad \mathcal{\tilde{F}}_{1}=-48\pi^{2}L^{3}\,u\,.
\end{equation}

\subsection{Field theory interpretation\label{FTHor-BaryonicFluxes}}

It would be very interesting to reproduce the entropy of the black hole horizons we found above through a field theory computation. The dual three-dimensional SCFTs for some specific $Y^{p,q}(\mathcal{B})$ have been studied in \cite{Martelli:2008si,Benini:2009qs,Benini:2011cma}. The dual of $M^{1,1,1}/\mathbb{Z}_k= Y^{2k,3k}(\mathbb{CP}^2)$ is a chiral theory and,
as we already mentioned, the large $N$ limit of the three-sphere partition function as well as  the topologically twisted index are not very well understood. The quiver for
$Q^{1,1,1}/\mathbb{Z}_k= Y^{k,k}(\mathbb{CP}^1 \times \mathbb{CP}^1)$ on the other hand is vector-like \cite{Benini:2009qs} and thus we have calculational control over the large $N$ limit of the localization calculations.
Any comparison of the black hole entropy for the near-horizon geometries discussed above with the field theory would require a proper understanding
of the cycles of the internal manifold and of the quantization of fluxes. The solution with no mesonic flavor and only baryonic magnetic flux presented in Section \ref{subsubsec:mag} should be the simplest case study. However, for reasons  that are still elusive, the fugacity and flux parameters associated with baryonic symmetries seem to disappear in the large $N$ limit of the twisted index, as is evident from the results in \cite{Hosseini:2016tor,Hosseini:2016ume}.
The same phenomenon has been noticed  for the $S^3$ free energy in \cite{Jafferis:2011zi}. These are all very interesting open problems that we leave for future work.

%%%%%%%%%%%%%%%%%%%%%%%%%%%%%%%%%%%%%
\section{Discussion}\label{sec:discussion}
%%%%%%%%%%%%%%%%%%%%%%%%%%%%%%%%%%%%%

In this paper we provided overwhelming evidence for the existence of a universal RG flow across dimensions between partially topologically twisted three-dimensional $\mathcal{N}=2$ SCFTs on $S^1\times \Sigma_\fg$ and one-dimensional quantum mechanical theories with two supercharges. This flow results in a universal relation between the topologically twisted index and the three-sphere partition function of the $\mathcal{N}=2$ SCFT, given by \eqref{eq:univintro}. If the three-dimensional SCFT admits a weakly-coupled supergravity dual description this universal RG flow has a simple holographic description in terms of a supersymmetric black hole solution of minimal four-dimensional gauged supergravity. This solution can then be embedded into string or M-theory in infinitely many distinct ways, which distinguish the different features of the $\mathcal{N}=2$ SCFTs.

There are a number of interesting open questions that stem from our results. The most obvious one is to understand in detail the plethora of AdS$_2$ solutions in M-theory presented in Section \ref{AdS2Hor}. The natural guess is that many of these solutions  correspond to twisted compactifications of three-dimensional $\mathcal{N}=2$ SCFTs with background flavor and baryonic fluxes. The three-dimensional theories should arise from M2-branes probing a $CY_4$ singularity in M-theory. The topological twist should then be realized by wrapping these branes on $\Sigma_\fg$ and the AdS$_2$ vacua we find should describe the low-energy behavior of this system upon backreaction. Clearly, these expectations await an explicit confirmation which hinges on a more detailed understanding of the global properties of the AdS$_2$ backgrounds in Section \ref{AdS2Hor}. A particularly puzzling feature is that in supergravity the background flux for baryonic $U(1)$ symmetries affects the details of the AdS$_2$ vacuum and thus the black hole entropy. On the other hand, it seems that such baryonic magnetic fluxes do not change the large $N$ limit of the topologically twisted index. It would be most interesting to resolve this apparent puzzle.

Another interesting avenue for generalizing our results is to incorporate electric charges into the supersymmetric black hole solutions we studied. While one can show that there is no dyonic generalization of the universal twist, it is clear that the more general supergravity solutions in Section \ref{AdS2Hor} should admit dyonic counterparts.\footnote{These dyonic solutions should be additional new examples of the type of dyonic black holes studied in the four-dimensional gauged supergravity literature. See for example \cite{Cacciatori:2009iz,DallAgata:2010ejj,Hristov:2010ri,Halmagyi:2013qoa} and reference thereof.} The entropy of these dyonic black hole near-horizon geometries can then be accounted for by a generalization of the topologically twisted index which incorporates both electric and magnetic flavor fluxes. This was pursued successfully in \cite{Benini:2016rke} for supersymetric dyonic black holes asymptotic to AdS$_4\times S^7$ which are dual to the ABJM theory on $S^1\times \Sigma_\fg$ deformed by magnetic fluxes on $\Sigma_\fg$ and Wilson lines along $S^1$.

We have limited the discussion in this work to the large $N$ limit of the topologically twisted index which, for theories with a holographic dual, is captured by the supergravity approximation of string or M-theory. It is of great interest to go beyond this limit, both in the field theory as well as in the gravitational analysis. On the field theory side it is natural to ask whether some remnant of the universal twist relation in \eqref{eq:univintro} survives beyond the large $N$ approximation. There might be reasons to be cautiously optimistic, given that similar universal relations were derived in \cite{Benini:2015bwz,BC} for the conformal anomaly coefficients of even-dimensional SCFTs related by RG flows across dimensions. On the gravity side the problem is equally challenging. One has to find subleading (in $N$) corrections to the entropy of the universal black hole presented in Section \ref{sec:A simple 4d black hole}. Perhaps the methods employed in \cite{Sen:2011ba} can be useful in finding these corrections. It is worth noting that a similar question was addressed successfully in \cite{Bhattacharyya:2012ye} for the subleading corrections to the $S^{3}$ partition function for three-dimensional $\mathcal{N}=2$ theories with an AdS$_4$ dual in M-theory. Resolving this question for the class of asymptotically AdS$_4$ black holes studied here is bound to teach us important lessons about holography and the quantum structure of black holes. 

\bigskip
\bigskip
\bigskip
\bigskip
\bigskip
\bigskip
\bigskip
%%%%%%%%%%%%%%%%%%%%%%%%%%%%%%%%%%%%%
\noindent \textbf{ Acknowledgements }
%%%%%%%%%%%%%%%%%%%%%%%%%%%%%%%%%%%%%
\medskip

\noindent We would like to thank Francesco Benini, Davide Cassani, Fri\dh rik Gautason, Adolfo Guarino, Seyed Morteza Hosseini, Dario Martelli, Noppadol Mekareeya, Krzysztof Pilch, James Sparks, and Phil Szepietowski for interesting discussions. FA is partially supported by INFN. The work of NB is supported in part by the starting grant BOF/STG/14/032 from KU Leuven and by an Odysseus grant G0F9516N from the FWO. The work of VSM is supported by a doctoral fellowship from the Fund for Scientific Research - Flanders (FWO) and in part by the ERC grant 616732-HoloQosmos. NB and VSM are also supported by the KU Leuven C1 grant ZKD1118 C16/16/005, by the Belgian Federal Science Policy Office through the Inter-University Attraction Pole P7/37, and by the COST Action MP1210 The String Theory Universe. PMC is supported by Nederlandse Organisatie voor Wetenschappelijk Onderzoek (NWO) via a Vidi grant. The work of PMC is part of the Delta ITP consortium, a program of the NWO that is funded by the Dutch Ministry of Education, Culture and Science (OCW). AZ is partially supported by the INFN and ERC-STG grant 637844-HBQFTNCER. NB and AZ would like to thank KIAS, Seoul for warm hospitality in the beginning stages of this project.

%%%%%%%%%%%%%%%%%%%%%%%%%%%%%%%%%%%%%
\begin{appendices}
%%%%%%%%%%%%%%%%%%%%%%%%%%%%%%%%%%%%%

%%%%%%%%%%%%%%%%%%%%%%%%%%%%%%%%%%%%%
\appendix

\section{The large $N$ index for massive IIA theories}
\label{app:fieldIIA}

The large $N$ limit of the topologically twisted index for three-dimensional $\mathcal{N}=2$ SCFTs with massive type IIA holographic duals has been derived in \cite{Hosseini:2016tor}. Here we present some more details on this construction.\footnote{Some of these results  arise from discussions with S. M. Hosseini and N. Mekareeya.} 
The Bethe potential $\cV(\Delta_I)$ is obtained by extremizing an auxiliary functional $\cV[\Delta_I,\rho(t),v(t)]$ with respect to the density $\rho(t)$ and  the distribution of eigenvalues $u(t) = N^{1/3}( i \, t+v(t))$ of the matrix model. The  functional $\cV[\Delta_I,\rho(t),v(t)]$ for a generic Yang-Mills-Chern-Simons theory with bi-fundamentals  in the limit $N\gg k_a$ is constructed as follows. There is a contribution
\begin{equation}
  {\quad \rule[-1.4em]{0pt}{3.4em} - i k_a N^{5/3} \int dt\, \rho(t)\, t\, v(t) + \frac{k_a}{2} N^{5/3} \int dt\, \rho(t)\, \left(t^2 - v(t)^2\right)\,}\;,
 \end{equation}
 for each Chern-Simons coupling $k_a$ and a contribution
  \begin{equation}
  {\quad \rule[-1.4em]{0pt}{3.4em} i  \, g_+\left(\Delta_{I}\right) \, N^{5/3} \int dt\, \frac{\rho(t)^2}{1-i v'(t)}}\;,
 \end{equation}
where $g_+(u) \equiv \frac{u^3}6 - \frac\pi2 u^2 + \frac{\pi^2}3 u$, for any  bi-fundamental field with chemical potential $\Delta_I$. These formulae are derived under the assumption that $0\le \Delta_I\le 2\pi$. Since the Bethe equations involve multivalued functions, one must treat this with care. Setting all  $k_a=k$, we find
\bea 
\cV[\Delta_I,\rho(t),v(t)] = i  N^{5/3} \int dt \left ( \rho(t)  \frac{G k}{2} ( - 2 t v(t) - i (t^2-v(t)^2) ) + \sum_I g_+(\Delta_I) \frac{\rho(t)^2}{1 - i v'(t)} \right )\, ,
\eea
where $G$ is the number of gauge groups. It is easy to extremize this functional with respect to $\rho(t)$ when 
\bea\label{assumption} 
\sum_{I\in a} \Delta_I =2\pi \, ,
\eea
for each term $W_a$ in the superpotential. 
The distribution of eigenvalues and the Bethe potential read
\bea\label{Bethe}  & \rho(t)= \frac{3\mu - 4 Gk t^2}{6\sqrt{3} (\sum_I g_+(\Delta_I))}\, ,\qquad  y(t)=-\frac{t}{\sqrt{3}} \, , \qquad \mu= (3 \sqrt{Gk} \sum_I g_+(\Delta_I))^{2/3} \;,  \\
&
\cV = i \frac{3^{13/6}}{20} \left ( 1- \frac{i}{\sqrt{3}}\right ) (\sum_I g_+(\Delta_I))^{2/3} k^{1/3} N^{5/3} 
\, .\eea
We can compare the result with the free energy on $S^3$ derived in \cite{Jafferis:2011zi,Fluder:2015eoa}. Although the set of rules  for constructing $\cV$ and $F_{S^3}$ seem different, 
the final result is the same. Indeed we find again relation (\ref{indexS3})
 \begin{equation}\label{indexS32}
-  \frac{2i }{\pi} \cV(\Delta_I) = F_{S^3} \left (\frac{\Delta_I}{\pi}\right ) \, . 
 \end{equation}
In order to check this relation, it is important to remember that, although $\Delta_I$ are parameterizing global symmetries, due to (\ref{assumption}), the quantities $\Delta_I/\pi$
behave for all calculational purposes as R-symmetry parameters. To this end it is convenient to introduce  trial central charges for the parent 4d quiver using the standard formulae \cite{Anselmi:1997am}
\bea 
a(\Delta_I/\pi)  = \frac{9}{32} \Tr R^3  -  \frac{3}{32}  \Tr R  \, , \qquad\qquad c(\Delta_I/\pi)  = \frac{1}{32}\left ( 9 \Tr R^3 - 5 \Tr R \right )  \, ,  
\eea
where we take a trace over all fermions  and we assign  R-charge $\Delta_I/\pi$ to the $I$-th chiral multiplet   and R-charge $1$ to the gauginos.  In the large $N$ limit, for theories with an AdS dual, we have  $c=a$, and therefore
\bea\label{TrR} \Tr R  = \left (G + \sum_I \left (\frac{\Delta_I}{\pi} -1\right ) \right ) N^2=0\, .\eea
Using this it follows that, in the large $N$ limit,
\bea  \sum_I g_+(\Delta_I) &=  \sum_I \frac{\pi^3}{6} \left [\left (\frac{\Delta_I}{\pi} -1\right )^3 -  \left (\frac{\Delta_I}{\pi} -1\right ) \right ]  \nonumber \\  &=
 \frac{\pi^3}{6} \left [\sum_I \left (\frac{\Delta_I}{\pi} -1\right )^3 + G\right ] 
 =  \frac{\pi^3}{6 N^2} \Tr R^3  =   \frac{16 \pi^3}{27 N^2} a(\Delta_I/\pi)  \, .\eea
Comparing this result with (3.30) in \cite{Fluder:2015eoa} we finally find 
 \begin{equation}\label{indexS32}
 %-  \frac{2i }{\pi} \cV(\Delta_I) = F_{S^3} \left (\frac{\Delta_I}{\pi}\right )  =    \frac{2^{5/3} 3^{1/6}\pi}{5} \left ( 1- \frac{i}{\sqrt{3}}\right )a(\Delta_I/\pi)^{2/3} (N k)^{1/3} \, . 
-  \frac{2i }{\pi} \cV(\Delta_I) = F_{S^3} \left (\frac{\Delta_I}{\pi}\right )  =    \frac{2^{5/3} 3^{1/6}\pi}{5} \left ( 1- \frac{i}{\sqrt{3}}\right )a(\Delta_I/\pi)^{2/3} (N k)^{1/3} \, . 
 \end{equation}
The index at large $N$ is obtained by combining (6.9) and (6.10) in \cite{Hosseini:2016tor} and reads\footnote{The introduction of $\fg$ is straightforward  - see Section 6 of \cite{Benini:2016hjo}.}
\bea\label{indexII}   \log Z &=& - N^{5/3}\left (  \frac{ \pi^2 G}{3} (1-\fg) +\sum_I  (\fn_I-1+\fg) g_+'(\Delta_I) \right ) \int dt\, \frac{\rho(t)^2}{1-i v'(t)} \,.
\eea
We want now to prove that
\bea
 \label{index theorem:attractor}
  \log Z  = i (1-\fg) \left ( \frac{2}{\pi} \cV \left( \Delta_I \right) +  \sum_{I} \left[ \left( \frac{\fn_I}{1-\fg} - \frac{\Delta_I}{\pi} \right) \frac{\partial\cV \left( \Delta_I \right)}{\partial \Delta_I} \right] \right)\,.
\eea

\begin{proof} First notice that we can consider all the $\Delta_I$ in  \eqref{index theorem:attractor} as independent variables and impose the constraints $\sum_{I \in a} \Delta_I  = 2 \pi$ only after differentiation. This is due to the form of the differential operator in \eqref{index theorem:attractor} and the topological twist constraint $\sum_{I \in a} \fn_I  = 2(1-\fg)$, as it is easy to check by an explicit computation.
To prove  \eqref{index theorem:attractor}, we promote the explicit factors of $\pi$ appearing in $g_+$ to a formal variable $\bm{\pi}$.
Notice that the ``on-shell'' Bethe potential $\mathcal{V}$, at large $N$, is a homogeneous function of $g_+$ and therefore of  $\Delta_I$ and $\bm{\pi}$, \ie\;
\begin{equation}
\mathcal{V}(\lambda \Delta_I, \lambda \bm{\pi}) = \lambda^2 \, \mathcal{V}(\Delta_I, \bm{\pi}) \, .
\end{equation}
Hence,
\begin{eqnarray}
  &&\frac{1}{\bm{\pi}} \left[ 2 \, \mathcal{V}(\Delta_I) -\sum_I  \Delta_I \frac{\partial \mathcal{V}(\Delta_I)}{\partial \Delta_I} \right]\, = \frac{\partial \mathcal{V}(\Delta_I, \bm{\pi})}{\partial \bm{\pi}}  =  i  N^{5/3} \int dt  \sum_I \left ( \frac23 \bm{\pi} \Delta_I -\frac12 \Delta_I ^2 \right )\frac{\rho(t)^2}{1 - i v'(t)}  \notag\\
&& = i  N^{5/3} \int dt  \sum_I \left [ - g_+'(\Delta_I) - \frac{\bm{\pi}^2}{3} \left(\frac{\Delta_I}{\bm{\pi}} -1\right)    \right ]\frac{\rho(t)^2}{1 - i v'(t)} \notag\\
&&= i  N^{5/3} \int dt \left [ -  \sum_I  g_+'(\Delta_I) +\frac{\bm{\pi}^2 G}{3} \right ]\frac{\rho(t)^2}{1 - i v'(t)}  \,,
\label{hom}
\end{eqnarray}
where we used (\ref{TrR}) and the fact that $\partial\cV/\partial \rho(t)=\partial\cV/\partial v(t)=0$ on shell. Similarly, 
\bea\label{der}
\sum_{I}  \fn_I  \frac{\partial\cV \left( \Delta_I \right)}{\partial \Delta_I} = i  N^{5/3} \int dt  \sum_I  \fn_I g_+'(\Delta_I) \frac{\rho(t)^2}{1 - i v'(t)}\,.
\eea
Multiplying (\ref{hom}) by $(1-\fg)$  and using (\ref{der}) and (\ref{indexII}) we see that indeed (\ref{index theorem:attractor}) holds.
\end{proof}

It is always possible to choose a parametrization for the $\Delta_I$, subject to the constraint in (\ref{assumption}), such that the trial  central charge $a$ is a homogeneous function
of degree three. In this case, (\ref{index theorem:attractor}) simplifies to
\bea
 \label{index theorem:attractor2}
  \log Z  = i   \sum_{I} \fn_I   \frac{\partial\cV \left( \Delta_I \right)}{\partial \Delta_I}  =  \hat{c}  (N k)^{1/3} \sum_{I} \fn_I   \frac{\partial a(\Delta_I/\pi)^{2/3} }{\partial \Delta_I}  \, ,
\eea
where we defined  $\hat{c}\equiv - \frac{2^{2/3} 3^{1/6}\pi^2}{5} ( 1- i/\sqrt{3})$. For example, for the  Chern-Simons theory in \cite{Jafferis:2011zi}, with three adjoints $\Phi_i$  and
superpotential $W= \Phi_1[\Phi_2,\Phi_3]$, with $\sum_I \Delta_I=2\pi$  we find 
\bea\label{IIAN=4} 
a(\Delta_I/\pi)=  \frac{27 N^2}{32 \pi^3} \Delta_1\Delta_2\Delta_3\, .
\eea

Similarly, for a $U(N)\times U(N)$
theory with equal Chern-Simons couplings $k$ based on the conifold quiver with superpotential $W=A_1 B_1 A_2 B_2 - A_1 B_2 A_2 B_1$ and the constraint $\sum_{I=1}^4 \Delta_I=2\pi$  we find 
\bea\label{IIAconifold} 
a(\Delta_I/\pi)=  \frac{27 N^2}{32 \pi^3} \sum_{I<J<K}\Delta_I\Delta_J\Delta_K\, .
\eea

The previous derivation is based on the assumption (\ref{assumption}). In principle, it is possible that there exist other extrema of $\cV$ in regions where $\sum_{I\in a} \Delta_I = 2\pi n$ with $n\ne 1$, and they  might contribute to the index.  For example,   for the quiver in \cite{Guarino:2015jca}, since  $0\le \Delta_I\le 2\pi$ we can have $\Delta_1+\Delta_2+\Delta_3 = 0, 2\pi, 4\pi, 6\pi$. It
is easy to see that the cases $0$ and $6\pi$ give singular solutions for $\rho(t)$ and $4\pi$ is related to $2\pi$ by the redefinition $\hat\Delta_i=2\pi -\Delta_I$. We have checked in many
models that, for the case of the universal twist,  where the fluxes are proportional to the exact R-charges, the other solutions are singular or related to  $\sum_{I\in a} \Delta_I = 2\pi$ by some discrete symmetry. However, we have no general proof of this fact. In general, one must alway check whether there are other saddle points for different values of the sum  $\sum_{I\in a} \Delta_I$.
For those other saddle points, the index theorem (\ref{index theorem:attractor})  does not hold and one needs to perform an explicit computation to check
whether they contribute or not. 

%%%%
\section{Uplift of universal solution to massive IIA} \label{sec:MIIA}

In this Appendix, we provide the details of the uplift of the universal solution \eqref{univ} to massive IIA supergravity using the  formulae of \cite{Guarino:2015vca,Varela:2015uca,Guarino:2015qaa}. The metric and dilaton read
\begin{equation}\label{eq:uplift1}
\begin{split}
d\hat{s}^2_{10} &= e^{\frac{2 \phi -\varphi }{8}} X^{\frac{1}{4}} \Delta_1^{\frac{1}{2}} \Delta_2^{\frac{1}{8}} \bigg( \widetilde{ds}^2_4 + g^{-2} e^{\varphi -2 \phi}  X^{-1} d\alpha^2 \\
& \quad + g^{-2} \sin ^2(\alpha ) \left(\Delta_1^{-1}ds^2_{\mathbb{CP}^2} + X^{-1} \Delta_2^{-1} \eta ^2\right) \bigg) , \\
e^{\hat \phi} &=  e^{\frac{1}{4}(6\phi+4\varphi)}X^{-1/2}\Delta_1^{-1}\Delta_2^{3/4} \,, \\
X &\equiv e^{2 \varphi } \chi ^2+1 \, , \\
\Delta_1 &\equiv e^{\varphi } \sin ^2(\alpha )+\cos ^2(\alpha ) \left(e^{2 \varphi } \chi ^2+1\right) e^{2 \phi -\varphi } \, , \\
\Delta_2 &\equiv e^{\varphi } \sin ^2(\alpha )+\cos ^2(\alpha ) e^{2 \phi -\varphi } \, , 
\end{split}
\end{equation}
where we write $\widetilde{ds}^2_4$ for the 4d metric, as it will turn out to be related by a rescaling to $ds_4^2$ from section \ref{sec:A simple 4d black hole} and ranges of the angles used is $0\leq\psi\leq2\pi$, $0\leq\alpha\leq\pi$. Furthermore, $\eta = d\psi + \sigma + g A^1$ and the standard Einstein metric on ${\mathbb{CP}^2}$ is given by
\begin{equation}\label{CP2}
\begin{split}
ds^2_{\mathbb{CP}^2} &= d\mu^2+\frac{\sin^2 \mu}{4} \left(\sigma_1^2+\sigma_2^2+\cos^2\mu \, \sigma_3^2 \right) \, , \\
\sigma_1 &\equiv \cos \alpha_3 d\alpha_1+\sin \alpha_1 \sin \alpha_3 d\alpha_2 \, , \\
\sigma_2 &\equiv \sin \alpha_3 d\alpha_1-\sin \alpha_1 \cos \alpha_3 d\alpha_2 \, , \\
\sigma_3 &\equiv d\alpha_3 + \cos \alpha_1 d\alpha_2 \, , \\
\sigma &\equiv \frac{\sin^2\mu}{2} \sigma_3 \, , \quad J = \frac{1}{2} d\sigma \, , \\
\end{split}
\end{equation}
where $J$ is the K\"ahler form on $\Bbb{CP}^{2}$, normalized such that $\int_{\mathbb{CP}^2} J \wedge J = 2 \, \text{vol}_{\mathbb{CP}^2} = \pi^2$.

Since we are interested in solutions that are asymptotic to the AdS$_{4}$ vacuum of the theory with $\mathcal{N}=2$ supersymmetry we set the scalars to (see Equation (7) in \cite{Guarino:2015jca} or  Table 3 in \cite{Guarino:2015qaa}):
\begin{equation}\label{eq:scalarIIA}
\chi = -\frac{1}{2} \left(\frac{m}{g}\right)^{\frac{1}{3}} \, , \quad e^{-\varphi} = \frac{\sqrt{3}}{2} \left(\frac{m}{g}\right)^{\frac{1}{3}} \, , \quad \zeta=\tilde{\zeta}=0 \, , \quad e^{-\phi} = \frac{1}{\sqrt{2}} \left( \frac{m}{g} \right)^\frac{1}{3} \, .
\end{equation}
We discuss the scalar $a$ below. For the solution of interest, the 4d metric and connection read 
\begin{equation}\label{eq:normA}
\widetilde{ds}_{4}^2 = \frac{m^{1/3}}{2 \sqrt{3} g^{7/3}} ds^2_4 \, ,\qquad  A^1= \frac{1}{3g} \frac{dx_1}{x_2} \,.
\end{equation}
%

%%%%%%%%%
\subsection{Determining the fluxes} \label{sec:AppUplift}
%%%%%%%%%

Here we show how to obtain the uplifted fluxes  \eqref{eq:uplift2A}. The massive IIA field strengths $\hat{F}_2, \hat H_3, \hat F_4$ are given in terms of their potentials by (see, e.g., Equation (A.4) in \cite{Guarino:2015vca})
\begin{equation}\label{eq:F234}
\begin{split}
\hat F_{(2)} &=  d \hat A_{(1)} + m \hat B_{(2)} \, , \\
\hat H_{(3)} &= d \hat B_{(2)} \, , \\
\hat F_{(4)} &= d \hat A_{(3)} + \hat A_{(1)} \wedge d \hat B_{(2)} + \frac{1}{2} m \hat B_{(2)} \wedge \hat B_{(2)} \, .
\end{split}
\end{equation}
The uplifted potentials $\hat A_{(1)}, \hat B_{(2)}, \hat A_{(3)}$  are given in terms of $SU(3)$-invariant fields in Equation (2.6) of \cite{Varela:2015uca}. Plugging these  into \eqref{eq:F234}, together with the values for the scalar fields \eqref{eq:scalarIIA}, gives
\eqs{ \nonumber
\hat F_2 &= -\frac{\sin (\alpha ) D a\wedge d \alpha}{g}+\frac{3 \sin (\alpha ) \left(\frac{m}{g}\right)^{2/3} \left(2 \sin ^2(\alpha )+\cos ^2(\alpha )\right) \eta \wedge d \alpha}{2 g \left(2 \sin ^2(\alpha )+3 \cos ^2(\alpha )\right)^2} -H_{(2)}^0 \cos (\alpha ) \\ \label{eq:F234nodual1}
&+\frac{H_{(2)}^1 \sin ^2(\alpha ) \cos (\alpha ) \left(\frac{m}{g}\right)^{2/3}}{2 \left(2 \sin ^2(\alpha )+3 \cos ^2(\alpha )\right)}-\frac{J \sin ^2(\alpha ) \cos (\alpha ) \left(\frac{m}{g}\right)^{2/3}}{g \left(\sin ^2(\alpha )+2 \cos ^2(\alpha )\right) \left(2 \sin ^2(\alpha )+3 \cos ^2(\alpha )\right)}\\ \label{eq:F234nodual2}
\hat H_3 &= \frac{\sin (\alpha ) \left(\frac{m}{g}\right)^{2/3} d \alpha \wedge H_{(2)}^1}{2 m}+\frac{2 \sin ^3(\alpha ) \left(\frac{m}{g}\right)^{2/3} d \alpha \wedge J}{g m \left(\sin ^2(\alpha )+2 \cos ^2(\alpha )\right)^2}+\frac{\sin (\alpha ) d \alpha \wedge \tilde{H}_{(2)0}}{g}-H_{(3)}^0 \cos (\alpha ) 
}
\begin{equation}\label{eq:F234nodual3}
\begin{split}
\hat F_4 &= \frac{a \sin (\alpha ) \cos (\alpha ) d \alpha \wedge H_{(3)}^0}{g}+\frac{\sin ^3(\alpha ) \cos (\alpha ) \left(\frac{m}{g}\right)^{2/3} D a\wedge d \alpha \wedge J}{g^2 m \left(\sin ^2(\alpha )+2 \cos ^2(\alpha )\right)}\\
&-\frac{\sin (\alpha ) \cos (\alpha ) \left(\frac{m}{g}\right)^{2/3} \eta \wedge d \alpha\wedge H_{(2)}^0}{2 g m}+\frac{\sin ^3(\alpha ) \cos (\alpha ) \left(\frac{m}{g}\right)^{1/3} \eta \wedge d \alpha \wedge H_{(2)}^1}{4 g^2 \left(2 \sin ^2(\alpha )+3 \cos ^2(\alpha )\right)}\\
&-\frac{3 \sin ^3(\alpha ) \cos (\alpha ) \left(\frac{m}{g}\right)^{1/3} \left(4 \sin ^2(\alpha )+5 \cos ^2(\alpha )\right) \eta \wedge d \alpha \wedge J}{2 g^3 \left(\sin ^2(\alpha )+2 \cos ^2(\alpha )\right) \left(2 \sin ^2(\alpha )+3 \cos ^2(\alpha )\right)}\\
&+\frac{\sin ^3(\alpha ) \cos (\alpha ) \left(\frac{m}{g}\right)^{2/3} \eta \wedge d \alpha \wedge \tilde{H}_{(2)0}}{2 g^2 \left(2 \sin ^2(\alpha )+3 \cos ^2(\alpha )\right)}-\frac{\sin (\alpha ) \cos (\alpha ) \eta \wedge d \alpha \wedge \tilde{H}_{(2)1}}{3 g^2}\\
&+\frac{\sin (\alpha ) \cos (\alpha ) d \alpha \wedge H_{(3)1}}{g}+\frac{\sin ^2(\alpha ) \cos ^2(\alpha ) \left(\frac{m}{g}\right)^{2/3} H_{(2)}^0\wedge J}{g m \left(\sin ^2(\alpha )+2 \cos ^2(\alpha )\right)}+\frac{\sin ^4(\alpha ) \left(\frac{m}{g}\right)^{1/3} H_{(2)}^1\wedge J}{2 g^2 \left(\sin ^2(\alpha )+2 \cos ^2(\alpha )\right)}\\
&+\frac{\sin ^4(\alpha ) \left(\frac{m}{g}\right)^{1/3} J\wedge J \left(2 \sin ^2(\alpha )+5 \cos ^2(\alpha )\right)}{2 g^3 \left(\sin ^2(\alpha )+2 \cos ^2(\alpha )\right)^2}+\frac{\sin ^2(\alpha ) J\wedge \tilde{H}_{(2)1}}{3 g^2}\\
&-\frac{\sin ^2(\alpha ) \cos ^2(\alpha ) \left(\frac{m}{g}\right)^{2/3} \eta \wedge H_{(3)}^0}{2 g \left(2 \sin ^2(\alpha )+3 \cos ^2(\alpha )\right)}+\frac{\sin ^2(\alpha ) \eta \wedge H_{(3)2}}{3 g}+H_{(4)}^0 \cos ^2(\alpha )+H_{(4)}^1 \sin ^2(\alpha ) \, ,
\end{split}
\end{equation}
where we replaced exterior derivatives of the invariant potentials $A^0,A^1,\tilde A_0,\tilde A_1, B^0,B_1,B_2,C^0,C^1$ by their field strengths, defined as,
\begin{align}
H^0_{(2)}&=dA^0+mB^0 \,, \quad & H^1_{(2)} &= d A^1 \, , \label{eq:H01}\\
\tilde H_{(2)0}&=d\tilde A_0 + g B^0 \,, \quad & \tilde H_{(2)1} &= d \tilde A_1-2 g B_2 \, , \label{eq:H01tilde}\\
H^0_{(3)} &= d B^0 \, , \quad &  H_{(3)2} &= d B_2 \, , \\
H_{(4)}^0 &= dC^0+H_{(2)}^0 \wedge B^0 - \frac{1}{2} m B^0 \wedge B^0 \,, \quad & H_{(4)}^1 &= dC^1-\frac{1}{3} H^1_{(2)} \wedge B_2 \, ,
\end{align}
\begin{equation}
\begin{split}
H_{(3)1} &=  d B_1-gA^0\wedge B^0+m\tilde A_0 \wedge B^0 + 2 g (C^1-C^0)\\
&+\frac{1}{2} \left(A^0\wedge d \tilde A_0 + \tilde A_0 \wedge d A^0 - \frac{1}{3} A^1 \wedge d \tilde A_1 - \frac{1}{3} \tilde A_1 \wedge d A^1\right) \, .
\end{split}
\end{equation}
The gauge potential $A^1$, and corresponding field strength $H^1_{(2)}$, are identified with the graviphoton $A$, and field strength $F$, of Section \ref{sec:A simple 4d black hole} up to a normalization, provided in \eqref{eq:normA}.
To fully determine the uplifted fluxes \eqref{eq:F234nodual1}-\eqref{eq:F234nodual3} we must determine the remaining $SU(3)$-invariant forms $H^0_{(2)},\tilde{H}_{(2)0}, \tilde{H}_{(2)1} , H_{(3)}^0, H_{(3)1}, H_{(3)2}, H_{(4)}^0, H_{(4)}^1$ in terms of the data of the minimal supergravity. A naive guess would be to set these to zero to obtain the minimal theory, but this is inconsistent with the duality relations of \cite{Guarino:2015qaa}. To determine the consistent result we note that setting 
\equ{\label{Da=0}
Da \equiv da+g A^0-m\tilde{A}_0=0
}
is consistent with the equations of motion of the Lagrangian (3.7) in \cite{Guarino:2015qaa}. Taking an exterior derivative of this equation implies $g \, dA^0=m \, d\tilde{A}_0$ and from the definitions \eqref{eq:H01} and \eqref{eq:H01tilde}, we find
\begin{equation}\label{eq:gHmH}
g H_{(2)}^0=m \tilde{H}_{(2)0} \, .
\end{equation}
%We can now use the duality relations to write the additional form fields in terms of the field strength $H^1_{(2)}=F$.
Then, using \eqref{Da=0} and \eqref{eq:gHmH} in the duality relations (3.17), (3.18), (3.19) in \cite{Guarino:2015qaa}, we obtain
%\eqs{\nonumber
%\tilde{H}_{(2)0} &= - \frac{1}{2}\left(\frac{g}{m}\right)^{1/3}(H_{(2)}^1-\sqrt{3} \star H_{(2)}^1) \,,  &&  \tilde{H}_{(2)1} =  - \frac{3}{2} \left(\frac{m}{g}\right)^{1/3} \left(H_{(2)}^1+\sqrt{3} \star H_{(2)}^1 \right) \,, \\ \label{HHs}
%H_{(3)}^0 &= H_{(3)1} = H_{(3)2}=0 \,, &&  H_{(4)}^0 = H_{(4)}^1= \frac{\text{vol}_4}{\sqrt{3}g^3} \left(\frac{m}{g}\right)^{1/3} \, .
%}
\eqs{\nonumber
\tilde{H}_{(2)0} &= - \frac{1}{2}\left(\frac{g}{m}\right)^{1/3}( H^1_{(2)}-\sqrt{3} \ast_{4}  H^1_{(2)}) \,,  &&  \tilde{H}_{(2)1} =  - \frac{3}{2} \left(\frac{m}{g}\right)^{1/3} \left( H^1_{(2)}+\sqrt{3} \ast_{4}  H^1_{(2)} \right) \,, \\ \label{HHs}
H_{(3)}^0 &= H_{(3)1} = H_{(3)2}=0 \,, &&  H_{(4)}^0 = H_{(4)}^1= \frac{\text{vol}_4}{\sqrt{3}g^3} \left(\frac{m}{g}\right)^{1/3} \, .
}
We see from here that it would have been inconsistent to set the fields $H^0_{(2)},\tilde H_{(2)0},\tilde H_{(2)1}, H^0_{(4)}, H^1_{(4)}$ to zero since they are related to $\text{vol}_4$ and $H^1_{(2)}$ by the duality relations. Finally, plugging \eqref{eq:gHmH} and \eqref{HHs} into the massive IIA field strengths \eqref{eq:F234nodual1} and \eqref{eq:F234nodual2} and \eqref{eq:F234nodual3} we obtain the result presented in \eqref{eq:uplift2A}. To minimize the possibility of mistakes introduced during this unwieldy uplifting procedure we have checked explicitly that the uplifted massive IIA black hole background is a solution of the ten-dimensional equations of motion. We now proceed to show that it preserves two supercharges.

%%%%%%%%%%%
\subsection{Proving supersymmetry} \label{sec:AppSUSYIIA}
%%%%%%%%%%%

Here we give the details establishing supersymmetry of the black hole solution \eqref{flowIIAmetric}, \eqref{eq:uplift2A} in massive IIA supergravity. It is useful to define the following basis of vielbeins
\begin{equation}\label{eq:vielbeinIIA}
\begin{split}
e^{1}&=e^{\lambda}L\left(\rho-\frac{1}{2\rho}\right)dt\,,\quad e^{2}=e^{\lambda}L \left(\rho-\frac{1}{2\rho}\right)^{-1}d\rho\,,\\
e^{3}&=e^{\lambda}L  \, \frac{\rho \, dx_{1}}{x_{2}}\,,\quad e^{4}=e^{\lambda}L \,\frac{\rho \, dx_{2}}{x_{2}}\,,\\
e^{5}&=\omega \,\sin(\alpha) \Delta_{1}^{-1/2}d\mu\,,\\
e^{6}&= \omega\,\sin(\alpha) \Delta_{1}^{-1/2} \frac{\sin \mu}{2}\sigma_{1}\,,\\
e^{7}&= \omega\,\sin(\alpha) \Delta_{1}^{-1/2} \frac{\sin \mu}{2}\sigma_{2}\,,\\
e^{8}&=\omega\,\sin(\alpha) \Delta_{1}^{-1/2} \frac{\sin \mu \cos \mu}{2}\sigma_{3}\,,\\
e^{9}&=\omega\,e^{\frac12(\varphi-2\phi)}X^{-1/2}d\alpha \,,\\
e^{10}&=\omega\, \sin(\alpha) X^{-1/2}\Delta_{2}^{-1/2}\eta\,,
\end{split}
\end{equation}
where
\begin{equation}\label{eq:lambdaomega}
\begin{split}
e^{2\lambda} &\equiv (\cos (2\alpha)+3)^{1/2}(\cos(2\alpha)+5)^{1/8}\,, \quad L^{2}\equiv \frac13 2^{-5/8}g^{-25/12}m^{1/12}\;,\\
\omega_{0} & \equiv \frac{2^{1/2} 3^{1/4} g^{1/6}}{m^{1/6}}\,, \quad \omega \equiv e^\lambda L\, \omega_0  \,.
\end{split}
\end{equation}
In our conventions $\Gamma_{11}=-\Gamma^{12345678910}$.

The general supersymmetry variations of massive IIA are written in \eqref{eq:var}. Using the basis \eqref{eq:vielbeinIIA} and plugging in the massive IIA forms \eqref{eq:uplift2A} we obtain a series of differential equations for $\epsilon$. Assuming that the spinor is independent of the coordinates $t,x_1,x_2,\mu,\alpha_1,\alpha_2,\alpha_3$ leads to a series of algebraic equations along these directions. It is useful to take linear combinations of \eqref{eq:var}, such as
\begin{equation}
\Gamma^1 \delta \psi_1 - \Gamma^2 \delta \psi_2 = N_0 \frac{\sqrt{3} \Gamma^2}{2\rho ^2} \left(1+2 \rho ^2+2\rho \left(1 - 2 \rho ^2 \right) \partial_{\rho }\right)\epsilon \, ,
\end{equation}
where $N_0^{-1} \equiv 2^{11/16} \sqrt[4]{\cos (2 \alpha )+3} \sqrt[16]{\cos (2 \alpha )+5} \sqrt[24]{\frac{m}{g^{25}}}$. 
The above equation is solved by taking the spinor be proportional to $\sqrt{\rho-\frac{1}{2\rho}}$.
We continue with
\begin{equation}
\Gamma^3 \delta \psi_3 - \Gamma^4 \delta \psi_4 = - \frac{2 N_0}{\sqrt{3} \rho } \Gamma^3 \left(\frac{3}{2} \Gamma^{34}+\partial_{\psi } \right) \epsilon \, ,
\end{equation}
which is solved by the $\psi$-dependence
\begin{equation}\label{eq:proj34}
\epsilon \propto e^{-\frac{3}{2}\Gamma^{34} \psi} \,. 
\end{equation}
Using this we find 
\begin{equation}
\Gamma^5 \delta \psi_5 - \Gamma^8 \delta \psi_8 \propto \left(\Gamma^{678} (3-\cos (2 \mu ))+2 \Gamma^5 \cos (2 \mu )+6 \Gamma^{348} \sin ^2(\mu ) \right) \epsilon \,.
\end{equation}
This is solved by imposing $\epsilon$ to be in the kernel of the following two projectors
\begin{equation}\label{eq:proj5678}
\Pi_1 \equiv \frac{1}{2}\left(1-\Gamma^{5678} \right) \,, \quad \Pi_{2} \equiv \frac{1}{2}\left(1+\Gamma^{3467} \right) \,.
\end{equation}
Using these projectors to replace $\Gamma^{34}$ and $\Gamma^{58}$ by $\Gamma^{67}$ we can write
\begin{equation}\label{eq:9}
\begin{split}
\frac{\delta \lambda}{2\sqrt{2}}-\Gamma^5 \delta \psi_5 &= N_0 \Big(-1 +\frac{2 \sqrt{3} \Gamma^{1267}}{\cos (2 \alpha )+5} +\frac{\Gamma^9 (22 \sin (2 \alpha )+\sin (4 \alpha )-96 \cot (\alpha ))}{\sqrt{2} (16 \cos (2 \alpha )+\cos (4 \alpha )+31)}\\
&+ \frac{4 \Gamma^{1210} \sin (\alpha )}{(\cos (2 \alpha )+3) \sqrt{\cos (2 \alpha )+5}}-\frac{\sqrt{3} \Gamma^{6710} (\cos (2 \alpha )+3) \csc (\alpha )}{2 \sqrt{\cos (2 \alpha )+5}}\Big) \epsilon \,.
\end{split}
\end{equation}
This equation can be interpreted as imposing a further projection condition (one can check that the combination on the right hand side of \eqref{eq:9} indeed squares to itself as a projector should). Now we can use the results so far to replace the $\Gamma^{34},\Gamma^{58},\Gamma^9$ gamma matrices. Doing so, we see that all variations vanish, except for $\delta \psi_{1,2,3,4,9}$.
The following combination is useful
\begin{equation}\label{eq:10}
\begin{split}
&\left(\Gamma^2 (\cos (2 \alpha )+5)+2 \sqrt{3} \Gamma^{167} \right) \left(\Gamma^1 \delta \psi_1+\Gamma^3 \delta \psi_3 \right) = N_0 \Big(4 \sqrt{3} \Gamma^{6710} \sin (\alpha ) \sqrt{\cos (2 \alpha )+5}\\
&+12 \Gamma^{167}+2 \sqrt{3} \Gamma^2 (\cos (2 \alpha )+5)-6 (\cos (2 \alpha )+3)\Big)\epsilon \, .
\end{split}
\end{equation}
The two equations \eqref{eq:9} and \eqref{eq:10} can be solved by imposing $\epsilon$ to be in the kernel of the following two projectors:
\begin{equation}\label{eq:pi34}
\begin{split}
\widehat \Pi_3 &\equiv \frac{1}{2}\left(1-\frac{2}{\cos (2 \alpha )+3}\Gamma^{134}+\frac{2 \sqrt{3} \cos ^2(\alpha )}{\cos (2 \alpha )+3} \Gamma^2-\frac{\sqrt{2} \sin (2 \alpha )}{\cos (2 \alpha )+3} \Gamma^9 \right) \,, \\
\widehat \Pi_4 &\equiv \frac{1}{2} \left(1 - \left(\frac{\sqrt{6} \cos (\alpha )}{\sqrt{\cos (2 \alpha )+5}}\Gamma^9+\frac{2 \sin (\alpha )}{\sqrt{\cos (2 \alpha )+5}}\Gamma^2 \right)\Gamma ^{6710}\right) \,.
\end{split}
\end{equation}
The reason for the notation with the ``hat'' in these two projectors will be made clear below.  Together with  \eqref{eq:proj5678}, we have now imposed four projectors, leaving $32\times 2^{-4}=2$ supercharges.
We note the projectors \eqref{eq:pi34} can be rewritten, inspired by the discussion in \cite{Pilch:2015dwa,Pilch:2015vha},  in a more insightful way as
\begin{equation}
\begin{split}
\widehat \Pi_3 &= \frac{1}{2}\left(1+\cos(\theta_2) \Gamma^{134}+\sin(\theta_2)\left(\cos(\theta_1)\Gamma^2+\sin(\theta_1)\Gamma^9\right)\right) \\
&=\mathcal{R}(\theta_1,\theta_2)\Pi^{}_3\mathcal{R}^{-1}\left(\theta_1,\theta_2\right)  \,, \\
\widehat \Pi_4 &= \frac{1}{2}\left(1-\left(\cos(\theta_1) \Gamma^{2}-\sin(\theta_1)\Gamma^9\right)\Gamma^{6710}\right) \\
&= \mathcal{R}(\theta_1,\theta_2)\Pi^{}_4 \mathcal{R}^{-1}(\theta_1,\theta_2)  \,,
\end{split}
\end{equation}
where $\Pi^{}_{3},\Pi^{}_{4}$ are the standard projectors 
\begin{equation}
\begin{split}
\Pi^{}_3 &\equiv \frac{1}{2}\left(1+\Gamma^{134}\right) \,, \\
\Pi^{}_4 &\equiv \frac{1}{2}\left(1+\Gamma^{67910}\right) \,,
\end{split}
\end{equation}
and $\mathcal{R}(\theta_1,\theta_2)$ is a rotation operator defined by
\eqsn{
\mathcal{R}(\theta_1,\theta_2) &\equiv \mathcal{R}_{29}(\theta_1) \mathcal{R}_{1234}(\theta_2) \,, \\
\mathcal{R}_{29}(\theta_1) &\equiv \left(\cos\left(\theta_1/2\right)-\sin\left(\theta_1/2\right)\Gamma^{29}\right) \,, \\
\mathcal{R}_{1234}(\theta_2) &\equiv \left(\cos\left(\theta_2/2\right)-\sin\left(\theta_2/2\right)\Gamma^{1234}\right) \,,
}
with the ``angles'' $\theta_{1}=\theta_1(\alpha),\theta_{2}=\theta_2(\alpha)$ defined as the following functions of $\alpha$
\begin{align}\label{eq:angles}
\cos(\theta_1/2) &\equiv \sqrt{\frac{\sqrt{\cos (2 \alpha )+5}-\sqrt{6} \cos (\alpha )}{2\sqrt{\cos (2 \alpha )+5}}} \,, & \sin(\theta_1/2) &\equiv \sqrt{\frac{\sqrt{6} \cos (\alpha )+\sqrt{\cos (2 \alpha )+5}}{2\sqrt{\cos (2 \alpha )+5}}} \,,\\
\cos(\theta_2/2) &\equiv \frac{\cos (\alpha )}{\sqrt{\cos (2 \alpha )+3}} \,, &\sin(\theta_2/2) &\equiv -\frac{1}{\sqrt{2}}\sqrt{\frac{\cos (2 \alpha )+5}{\cos (2 \alpha )+3}} \,.
\end{align}
The spinor can thus be written as
\begin{equation}
\epsilon=\mathcal{R}(\theta_1,\theta_2)\tilde{\epsilon}\,,
\end{equation}
where $\tilde \epsilon$ is a spinor  in the kernel of the unrotated projectors $\Pi_1,\Pi_2,\Pi^{}_3,\Pi^{}_4$.
The only equation remaining at this point is $\delta \psi_9=0$, which can be solved by taking $\tilde \epsilon$ to be proportional to the function $e^{\lambda/2}$ defined in \eqref{eq:lambdaomega}.

We thus find that our massive type IIA black hole solution is indeed supersymmetric, with the following explicit Killing spinor
\begin{equation}
\epsilon = \sqrt{\rho-\frac{1}{2\rho}} e^{\lambda/2} e^{\frac{3}{2}\Gamma^{67}\psi}\mathcal{R}(\theta_1,\theta_2) \eta \, ,
\end{equation}
where $\eta$ is a constant spinor in the kernel of the four projectors $\Pi_1,\Pi_2,\Pi^{}_3,\Pi^{}_4$ and hence there are a total of 2 supercharges preserved. Note that the factor  $\sqrt{\rho-\frac{1}{2\rho}} e^{\lambda/2}$ is precisely proportional to $|g_{tt}|^{1/4}$, where  $g_{tt}$ is the time-time component of the ten-dimensional metric in \eqref{eq:uplift1}. 

%%%%%%%%%%%%%%%%%%%%%%%%%%%%%%%%%%%%%

\section{Details on the construction of the 11D solutions}
\label{Appendix11D}

In this appendix we give the details of the derivation of the family
of solutions of eleven dimensional supergravity described in Section
$\ref{AdS2Hor}$. Before starting, it is convenient to choose an elfbein:
\begin{gather}
e^{0}=\frac{f_{1}}{z}dt\,,\quad e^{1}=\frac{f_{1}}{z}dz\,,\nonumber \\
e^{2}=f_{2}e^{h}dx_{1}\,,\quad e^{3}=f_{2}e^{h}dx_{2}\,,\nonumber \\
e^{4}=f_{3}dy\,,\quad e^{5}=f_{4}D\beta\,,\label{AdS2Hor_elfbein}\\
e^{6}=f_{5}E^{1}\,,\quad e^{7}=f_{5}E^{2}\,,\quad e^{8}=f_{5}E^{3}\,,\quad e^{9}=f_{5}E^{4}\,,\nonumber \\
e^{10}=f_{7}D\psi\,,\nonumber 
\end{gather}
where $(t,z)$ are the Poincar\'{e} coordinates for AdS$_{2}$ and
$E^{1,2,3,4}$ is a vierbein on the K\"ahler manifold $\mathcal{B}$ such that
\begin{equation}
J_{\mathcal{B}}=E^{1}\wedge E^{2}+E^{3}\wedge E^{4}\,.\label{AdS2Hor_BKaehlerForm}
\end{equation}
We can thus rewrite the 4-form as
\begin{eqnarray*}
G_{4} & = & e^{0}\wedge e^{1}\wedge\left(g_{1}\,e^{2}\wedge e^{3}+g_{2}\,e^{4}\wedge e^{5}+g_{3}\,e^{5}\wedge e^{10}+g_{4}\,e^{4}\wedge e^{10}+g_{5}\,\left(e^{6}\wedge e^{7}+e^{8}\wedge e^{9}\right)\right)\\
 &  & +e^{2}\wedge e^{3}\wedge\left(i_{1}\,e^{4}\wedge e^{5}+i_{2}\,(e^{6}\wedge e^{7}+e^{8}\wedge e^{9})+i_{4}\,e^{5}\wedge e^{10}+i_{5}\,e^{4}\wedge e^{10}\right)\\
 &  & +\left(e^{6}\wedge e^{7}+e^{8}\wedge e^{9}\right)\wedge\left(l_{1}\,e^{4}\wedge e^{5}+l_{2}\,\left(e^{6}\wedge e^{7}+e^{8}\wedge e^{9}\right)+l_{3}\,e^{5}\wedge e^{10}+l_{4}\,e^{4}\wedge e^{10}\right)\,.
\end{eqnarray*}

\subsection{BPS equations}
Imposing the vanishing of the gravitino variations in eleven-dimensional supergravity gives the following Killing spinor equation:
\begin{equation}
\delta \psi_{\mu}=\nabla_{\mu}\varepsilon+\frac{G_{\alpha\beta\gamma\delta}^{4}}{288}\left(\Gamma_{\mu}^{\;\alpha\beta\gamma\delta}-8\delta_{\mu}^{\alpha}\Gamma^{\beta\gamma\delta}\right)\varepsilon=0\,.
\label{AdS2Hor_BPSeq}
\end{equation}
Motivated by the symmetry of the problem and the underlying M2-brane interpretation we impose the following projection conditions on $\varepsilon$:
\begin{equation}
\Gamma^{23}\varepsilon=\Gamma^{45}\varepsilon=\Gamma^{67}\varepsilon=\Gamma^{89}\varepsilon=\Gamma^{1\,10}\varepsilon=i\varepsilon\,.
\label{AdS2Hor_BPSProjections}
\end{equation}
Equation $\eqref{AdS2Hor_BPSeq}$ leads to the system of
differential equations:\footnote{Here it is convenient to anticipate one of results from the
Bianchi identity for $G_{4}$, namely $g_{3},i_{4},l_{3}=0$.}
\begin{align}
\frac{f_{1}'}{2f_{1}f_{3}}+\frac{g_{4}}{6} & =0\,,\label{AdS2Hor_BPSDiffEq1}\\
\frac{f_{2}'}{f_{2}f_{3}}-\frac{g_{4}}{6}-\frac{1}{2}\frac{cf_{4}}{f_{2}^{2}} & =0\,,\label{AdS2Hor_BPSDiffEq2}\\
\frac{f_{5}'}{f_{5}f_{3}}-\frac{g_{4}}{6}-\frac{f_{4}b}{2f_{5}^{2}} & =0\,,\label{AdS2Hor_BPSDiffEq3}\\
\frac{1}{2}\frac{f_{7}'}{f_{7}f_{3}}+\frac{g_{4}}{6} & =0\,,\label{AdS2Hor_BPSDiffEq4}\\
-\frac{3}{f_{1}}+g_{1}+g_{2}+2g_{5} & =0\,,\label{AdS2Hor_BPSgEq1}\\
\frac{1}{4}\frac{f_{7}\left(\tilde{c}-f_{8}c\right)}{f_{2}^{2}}-\frac{-2g_{1}+g_{2}+2g_{5}}{12} & =0\,,\label{AdS2Hor_BPSgEq2}\\
-\frac{1}{4}\frac{f_{7}f_{8}'}{f_{3}f_{4}}-\frac{g_{1}-2g_{2}+2g_{5}}{12} & =0\,,\label{AdS2Hor_BPSgEq3}\\
\frac{1}{4}\frac{f_{7}\left(a-f_{8}b\right)}{f_{5}^{2}}-\frac{g_{1}+g_{2}-g_{5}}{12} & =0\,,\label{AdS2Hor_BPSgEq4}
\end{align}
some constraints on the $i_{n}$ and $l_{n}$ coefficients,
\begin{equation}
i_{1}=-2i_{2}\,,\quad l_{1}=i_{2}\,,\quad l_{2}=-i_{2}\,,\quad i_{5}=l_{4}=0\,,\label{AdS2Hor_ilEqs}
\end{equation}
and a set of PDEs for $\varepsilon$:
\begin{alignat}{1}
\partial_{t}\varepsilon & =0\,,\label{AdS2Hor_BPS-PDE1}\\
z\partial_{z}\varepsilon+\frac{1}{2}\varepsilon & =0\,,\label{AdS2Hor_BPS-PDE2}\\
\frac{1}{f_{3}}\partial_{y}\varepsilon+\frac{g_{4}}{6}\varepsilon & =0\,,\label{AdS2Hor_BPS-PDE3}\\
\partial_{\beta}\varepsilon+f_{8}\partial_{\psi}\varepsilon-\frac{i}{2}f_{4}\left(\frac{f_{4}'}{f_{3}f_{4}}-\frac{g_{4}}{6}+\frac{cf_{4}}{2f_{2}^{2}}+\frac{f_{4}b}{f_{5}^{2}}\right)\varepsilon & =0\,,\label{AdS2Hor_BPS-PDE4}\\
\left(\partial_{x_{1}}-cA_{1}\partial_{\beta}-\tilde{c}A_{1}\partial_{\psi}\pm\frac{i}{2}\partial_{x_{2}}h\right)\varepsilon & =0\,,\label{AdS2Hor_BPS-PDE5}\\
\left(\partial_{x_{2}}-cA_{2}\partial_{\beta}-\tilde{c}A_{2}\partial_{\psi}\mp\frac{i}{2}\partial_{x_{1}}h\right)\varepsilon & =0\,,\label{AdS2Hor_BPS-PDE6}\\
\left(\nabla_{\mathcal{B}}^{i}-A_{\mathcal{B}}^{i}\left(a\partial_{\psi}+b\partial_{\beta}\right)\right)\varepsilon & =0\,,\label{AdS2Hor_BPS-PDE7}
\end{alignat}
\begin{equation}
\frac{4}{f_{7}}\partial_{\psi}\varepsilon-i\frac{1}{4}\left(\frac{f_{7}\left(\tilde{c}-f_{8}c\right)}{f_{2}^{2}}+2\frac{f_{7}\left(a-f_{8}b\right)}{f_{5}^{2}}-\frac{f_{7}f_{8}'}{f_{3}f_{4}}+\frac{g_{1}+g_{2}+2g_{5}}{3}\right)\varepsilon=0\,,\label{AdS2Hor_BPS-PDE9}
\end{equation}
where $\nabla_{\mathcal{B}}^{i}$ denotes the covariant derivative
on $\mathcal{B}$ and $i=1,2,3,4$ spans its four real coordinates.

Let us now solve for the dependence of $\varepsilon$ on the various
coordinates. From $\eqref{AdS2Hor_BPS-PDE1}$
we deduce that it is constant in time. We also assume that $\partial_{x_{1}}\varepsilon=\partial_{x_{2}}\varepsilon=0$
and that $\varepsilon$ does not depend on the coordinates of $\mathcal{B}$,
checking \textit{a posteriori} the consistency of such assumption.

The coefficients $f_{1}$ and $f_{7}$ are related by $\eqref{AdS2Hor_BPSDiffEq1}$
and $\eqref{AdS2Hor_BPSDiffEq4}$, which can be combined to obtain
\begin{equation}
\frac{f_{1}'}{f_{1}}=\frac{f_{7}'}{f_{7}}\,,\label{AdS2Hor_f1f7DiffEq}
\end{equation}
and thus
\begin{equation}
f_{7}=2\omega_{1}f_{1}\,,\label{AdS2Hor_f1f7}
\end{equation}
with $\omega_{1}$ an integration constant.

Then, adding $\eqref{AdS2Hor_BPSgEq2}$, $\eqref{AdS2Hor_BPSgEq3}$
and $\eqref{AdS2Hor_BPSgEq4}$ and using $\eqref{AdS2Hor_BPSgEq1}$
and $\eqref{AdS2Hor_f1f7DiffEq}$, $\eqref{AdS2Hor_BPS-PDE9}$ leads to
\begin{equation}
\partial_{\psi}\varepsilon=i\omega_{1}\varepsilon\,.\label{AdS2Hor_epsilonPsiDep}
\end{equation}
Then, from equations $\eqref{AdS2Hor_BPS-PDE5}$ and $\eqref{AdS2Hor_BPS-PDE6}$
we find
\[
\left(-ic\partial_{\beta}\varepsilon+\tilde{c}\omega_{1}\varepsilon\right)dA=-\frac{1}{2}\left(\partial_{x_{1}}^{2}h+\partial_{x_{2}}^{2}h\right)dx_{1}\wedge dx_{2}\,.
\]
Using the constant curvature metric \eqref{defh}, the combination on the r.h.s of this equation is constant and proportional to
$\kappa$, the normalized curvature of the Riemann surface $\Sigma_{\mathfrak{g}}$,
\[
\kappa=\begin{cases}
1 & \mathrm{for\;}\mathfrak{g}=0\\
0 & \mathrm{for\;}\mathfrak{g}=1\\
-1 & \mathrm{for\;}\mathfrak{g}>1
\end{cases}\,.
\]
Thus, we find the dependence of $\varepsilon$  on $\beta$ is given by
\begin{equation}
\partial_{\beta}\varepsilon\equiv i\omega_{2}\,,\label{AdS2Hor_epsilonBetaDep}
\end{equation}
with $\omega_{2}$ a constant satisfying 
\begin{equation}
c\omega_{2}+\tilde{c}\omega_{1}=\frac{\kappa}{2}\,.\label{AdS2Hor_cCostraint}
\end{equation}
One can think of this algebraic constraint as the supergravity implementation of the topological twist condition in the dual 3d $\mathcal{N}=2$ SCFT.

Equations $\eqref{AdS2Hor_BPS-PDE3}$ and $\eqref{AdS2Hor_BPSDiffEq1}$
allow us to determine also the dependence on $y$:
\begin{equation}
\partial_{y}\varepsilon=\frac{f_{1}'}{2f_{1}}\varepsilon=\frac{d\left(\sqrt{f_{1}}\right)}{dy}\varepsilon\,,\label{AdS2Hor_epsilonYDep}
\end{equation}
while $\eqref{AdS2Hor_BPS-PDE2}$ the one on $z$:
\[
\varepsilon\propto\frac{1}{\sqrt{z}}\,.
\]
Combining this result with $\eqref{AdS2Hor_epsilonPsiDep}$, $\eqref{AdS2Hor_epsilonBetaDep}$
and $\eqref{AdS2Hor_epsilonYDep}$ we see that
\begin{equation}
\varepsilon=e^{i\left(\omega_{2}\beta+\omega_{1}\psi\right)}\sqrt{\frac{f_{1}}{z}}\varepsilon_{0}\,,\label{AdS2Hor_epsilonFull}
\end{equation}
with $\varepsilon_{0}$ a constant spinor. 

Let us now turn to condition $\eqref{AdS2Hor_BPS-PDE7}$. Referring
to $\eqref{AdS2Hor_elfbein}$ and $\eqref{AdS2Hor_BKaehlerForm}$
for the indices of the coordinates on $\mathcal{B}$, for both $\mathbb{CP}^{1}\times\mathbb{CP}^{1}$
and $\mathbb{CP}^{2}$, equipped with the standard Fubini-Study metric, we have that 
\begin{equation}
\omega_{\mathcal{B}}^{13}=\omega_{\mathcal{B}}^{24}=\omega_{\mathcal{B}}^{23}=\omega_{\mathcal{B}}^{41}\,,\label{AdS2Hor_BConnectionSym}
\end{equation}
where $\omega_{\mathcal{B}}$ is the spin connection in $\mathcal{B}$.
Therefore equation $\eqref{AdS2Hor_BPS-PDE7}$:
\begin{equation}
\frac{1}{4}\omega_{{\cal B}}^{jk}\Gamma^{(j+5)(k+5)}\varepsilon=iA_{{\cal B}}\left(a\omega_{1}+b\omega_{2}\right)\varepsilon\label{AdS2Hor_abConstraintProof}
\end{equation}
can be rewritten by $\eqref{AdS2Hor_BConnectionSym}$ as
\[
\frac{1}{2}\left(\omega_{{\cal B}}^{12}+\omega_{{\cal B}}^{34}\right)=A_{{\cal B}}\left(a\omega_{1}+b\omega_{2}\right)\,,
\]
since other combinations, such as
\[
\left(\omega_{\mathcal{B}}^{31}\Gamma^{86}+\omega_{\mathcal{B}}^{24}\Gamma^{79}\right)\varepsilon=\left(\omega_{\mathcal{B}}^{31}+\omega_{\mathcal{B}}^{24}\right)\Gamma^{78}\varepsilon=0\,,
\]
vanish identically. On the other hand, the fact that $\mathcal{B}$
is Einstein, implies that its Ricci 2-form is proportional to its
K\"{a}hler form by a constant $2q$:
\begin{equation}
\frac{1}{2}\left(d\omega_{\mathcal{B}\;\; j}^{\;i}+\omega_{\mathcal{B}\;k}^{i}\wedge\omega_{\mathcal{B}\;\; j}^{\;k}\right)J_{\mathcal{B}\,i}^{\;\;\;\; j}=2qJ_{\mathcal{B}}\,.\label{AdS2Hor_KECondition}
\end{equation}
We notice now that when one computes the only non-vanishing contributions
to $\eqref{AdS2Hor_KECondition}$ from the Riemann 2-form, namely
the components $(1,2)$ and $(3,4)$, the quadratic part vanishes
because of $\eqref{AdS2Hor_BConnectionSym}$. For instance one finds
\[
\omega_{\mathcal{B}\;3}^{1}\wedge\omega_{\mathcal{B}}^{32}+\omega_{\mathcal{B}\;4}^{1}\wedge\omega_{\mathcal{B}}^{42}=0\,.
\]
Therefore, taking the differential in $\eqref{AdS2Hor_abConstraintProof}$,
we can replace the first member with the second of $\eqref{AdS2Hor_KECondition}$
and, by $\eqref{AdS2Hor_ConnectionRels}$, arrive at the following constraint: 
\begin{equation}
b\omega_{2}+a\omega_{1}=q\,.\label{AdS2Hor_abConstraint}
\end{equation}
From equations $\eqref{AdS2Hor_BPSgEq1}$,$\eqref{AdS2Hor_BPSgEq2}$,$\eqref{AdS2Hor_BPSgEq3}$,
$\eqref{AdS2Hor_BPSgEq4}$ and $\eqref{AdS2Hor_BPSDiffEq1}$, we can
express the functions $g_{i}$ in terms of the metric funtions:
\begin{equation}
\begin{split}g_{1} & =-\frac{f_{7}\left(\tilde{c}-f_{8}c\right)}{f_{2}^{2}}+\frac{1}{f_{1}}\,,\\
g_{2} & =\frac{f_{7}f_{8}'}{f_{3}f_{4}}+\frac{1}{f_{1}}\,,\\
g_{4} & =-3\frac{f_{1}'}{f_{1}f_{3}}\,,\\
g_{5} & =-\frac{f_{7}\left(a-f_{8}b\right)}{f_{5}^{2}}+\frac{1}{f_{1}}\,,
\end{split}
\label{AdS2Hor_gSolution}
\end{equation}
and find the constraint
\begin{equation}
\frac{f_{7}\left(\tilde{c}-f_{8}c\right)}{f_{2}^{2}}+2\frac{f_{7}\left(a-f_{8}b\right)}{f_{5}^{2}}-\frac{f_{7}f_{8}'}{f_{3}f_{4}}=\frac{1}{f_{1}}\,,\label{AdS2Hor_f1Eq}
\end{equation}
which then can be solved by using $\eqref{AdS2Hor_f1f7}$ to eliminate $f_{7}$.
This results in the relation,
\begin{equation}
\frac{-f_{2}^{2}f_{5}^{4}+2\omega_{1}\left(\tilde{c}-f_{8}c\right)f_{1}^{2}f_{5}^{4}+4\omega_{1}\left(a-f_{8}b\right)f_{2}^{2}f_{1}^{2}f_{5}^{2}-2\frac{\omega_{1}}{K}f_{1}^{3}f_{2}^{2}f_{5}^{4}f_{8}'}{f_{2}^{2}f_{5}^{4}f_{1}^{2}}=0\,.\label{AdS2Hor_F1SolProof1}
\end{equation}
We note that $f_{3}$ is in fact redundant, as it can be set to an arbitrary function  by redefinitions of the coordinate $y$. A convenient choice is such that 
\begin{equation}
f_{1}(y)f_{3}(y)f_{4}(y)=K\,,\label{AdS2Hor_f134KConstraint}
\end{equation}
with $K$ an arbitrary constant; we consider this as an equation determining $f_{4}$. Taking also \eqref{AdS2Hor_f1f7} into account, we are left with  $f_{1,2,3,5,8}$ as five independent functions. It is convenient to trade these for $F_{1,2,3,5,6}$, defined by:
\begin{equation}
F_{1}\equiv f_{2}^{2}f_{5}^{4},\quad F_{2}\equiv f_{1}f_{2}^{2},\quad F_{3}\equiv f_{1}f_{5}^{2},\quad F_{5}\equiv K\frac{f_{1}^{2}f_{2}^{2}f_{5}^{4}}{f_{3}^{2}},\quad F_{6}\equiv f_{8}f_{1}^{3}f_{2}^{2}f_{5}^{4}\,.
\label{AdS2Hor_FBasis}
\end{equation}
In terms of this new basis, equations $\eqref{AdS2Hor_BPSDiffEq2}$ and $\eqref{AdS2Hor_BPSDiffEq3}$ simplify to 
\begin{equation}
F_{2}'=cK,\quad F_{3}'=bK\,,\label{AdS2Hor_KEqs}
\end{equation}
and thus
\equ{\label{F2F3app}
F_{2}=c K\,y+S_{2},\quad F_{3}=b K\,y+S_{3}\,,
}
with $S_{2}$ and $S_{3}$ integration constants. 

Furthermore, the complicated equation  $\eqref{AdS2Hor_F1SolProof1}$ becomes
\begin{equation}
F_{1}=2\omega_{1}\left(-\frac{F_{6}'}{K}+\tilde{c}F_{3}^{2}+2aF_{2}F_{3}\right).\label{AdS2Hor_SolutionF1}
\end{equation}
Finally, $F_{6}$ can be found through $\eqref{AdS2Hor_BPS-PDE4}$, which,
after using $\eqref{AdS2Hor_BPSDiffEq1}$, $\eqref{AdS2Hor_epsilonBetaDep}$,
$\eqref{AdS2Hor_epsilonPsiDep}$ and $\eqref{AdS2Hor_f134KConstraint}$,
reads
\begin{equation}
F_{6}=\frac{F_{5}'-4\omega_{2}F_{2}F_{3}^{2}}{4\omega_{1}}\,.\label{AdS2Hor_Solution-F6}
\end{equation}
Thus, at this point the solution is completely controlled by the single function $F_{5}$, which is determined by the Bianchi identity for $G_{4}$, as we discuss next.

%%%%%%%%%%%%%%%
\subsection{4-form equations}
\label{4-form equations app}
%%%%%%%%%%%%%%%

The equation of motion for the 4-form,
\[
d\ast_{11}\,G_{4}=0\,,
\]
implies the following relations:
\begin{align}
g_{4}i_{2}f_{1}^{2}f_{7}f_{3}f_{5}^{4}+\left(i_{2}f_{7}f_{1}^{2}f_{5}^{4}\right)'+bi_{2}f_{3}f_{4}f_{7}f_{1}^{2}f_{5}^{2} & =0\,,\label{AdS2Hor_4FormEoM1}\\
\left(i_{2}f_{7}f_{1}^{2}f_{5}^{2}f_{2}^{2}\right)'-ci_{2}f_{3}f_{4}f_{7}f_{1}^{2}f_{5}^{2}+2bi_{2}f_{3}f_{4}f_{2}^{2}f_{7}f_{1}^{2}+g_{4}i_{2}f_{1}^{2}f_{7}f_{3}f_{2}^{2}f_{5}^{2} & =0\,,\label{AdS2Hor_4FormEoM2}\\
i_{2}f_{7}f_{1}^{2}f_{2}^{2}f_{5}^{2}\left(a-bf_{8}\right)-i_{2}f_{7}f_{1}^{2}f_{5}^{4}\left(\tilde{c}-cf_{8}\right)+\left(g_{1}i_{2}-g_{5}i_{2}\right)f_{1}^{2}f_{2}^{2}f_{5}^{4} & =0\,,\label{AdS2Hor_4FormEoM3}\\
i_{2}f_{8}'f_{7}f_{1}^{2}f_{5}^{2}+i_{2}f_{3}f_{4}f_{7}f_{1}^{2}f_{5}^{2}\left(a-bf_{8}\right)-\left(g_{2}i_{2}-g_{5}i_{2}\right)f_{1}^{2}f_{3}f_{4}f_{5}^{4} & =0\,,\label{AdS2Hor_4FormEoM4}\\
2i_{2}f_{3}f_{4}f_{2}^{2}f_{7}f_{1}^{2}\left(a-bf_{8}\right)+i_{2}f_{8}'f_{7}f_{1}^{2}f_{5}^{2}f_{2}^{2}-i_{2}f_{3}f_{4}f_{7}f_{1}^{2}f_{5}^{2}\left(\tilde{c}-cf_{8}\right)\nonumber \\
-\left(g_{2}i_{2}-2g_{5}i_{2}+g_{1}i_{2}\right)f_{1}^{2}f_{2}^{2}f_{3}f_{4}f_{5}^{2} & =0\,,\label{AdS2Hor_4FormEoM5}\\
\left(g_{2}f_{2}^{2}f_{5}^{4}f_{7}\right)'-cg_{1}f_{3}f_{4}f_{5}^{2}f_{6}^{2}f_{7}-2bg_{5}f_{3}f_{4}f_{2}^{2}f_{5}^{2}f_{7} & =0\,,\label{AdS2Hor_4FormEoM6}\\
\tilde{c}g_{1}f_{3}f_{4}f_{5}^{4}f_{7}-\left(f_{8}g_{2}f_{2}^{2}f_{5}^{4}f_{7}\right)'+2ag_{5}f_{3}f_{4}f_{5}^{2}f_{2}^{2}f_{7}-\left(g_{4}f_{2}^{2}f_{4}f_{5}^{4}\right)'-6f_{2}^{2}f_{3}f_{4}f_{5}^{4}i_{2}^{2} & =0\,.\label{AdS2Hor_4FormEoM7}
\end{align}
It turns out that after imposing the BPS equations studied above, all  these equations are satisfied automatically. The first two are actually equivalent to the conditions
$\eqref{AdS2Hor_Bianchi3}$ and $\eqref{AdS2Hor_Bianchi4}$ implied
by the Bianchi identity that we examine later, after inserting $\eqref{AdS2Hor_BPSDiffEq1}$
and $\eqref{AdS2Hor_f1f7}$. Equations $\eqref{AdS2Hor_4FormEoM3}$,
$\eqref{AdS2Hor_4FormEoM4}$, $\eqref{AdS2Hor_4FormEoM5}$ are identically
satisfied when we plug $\eqref{AdS2Hor_gSolution}$ and $\eqref{AdS2Hor_ilEqs}$
in. The same happens to $\eqref{AdS2Hor_4FormEoM6}$ and $\eqref{AdS2Hor_4FormEoM7}$
using $\eqref{AdS2Hor_f134KConstraint}$ and $\eqref{AdS2Hor_KEqs}$.

On the other hand, the Bianchi identity,
\[
dG_{4}=0\,,
\]
provides a wealth of information: it implies, as mentioned before,
that $g_{3}=i_{4}=l_{3}=0$ and  the four additional equations
\begin{align}
-bf_{1}^{2}f_{3}f_{4}g_{2}-f_{1}^{2}f_{3}f_{7}g_{4}\left(a-bf_{8}\right)+\left(f_{1}^{2}f_{5}^{2}g_{5}\right)' & =0\,,\label{AdS2Hor_Bianchi1}\\
\left(f_{1}^{2}f_{2}^{2}g_{1}\right)'-c\left(f_{1}^{2}f_{3}f_{4}g_{2}\right)-f_{1}^{2}f_{3}f_{7}g_{4}\left(\tilde{c}-cf_{8}\right) & =0\,,\label{AdS2Hor_Bianchi2}\\
-bf_{2}^{2}f_{3}f_{4}i_{1}-cf_{5}^{2}f_{3}f_{4}l_{1}+\left(f_{2}^{2}f_{5}^{2}i_{2}\right)'-f_{2}^{2}f_{3}f_{7}i_{5}\left(a-bf_{8}\right)-f_{5}^{2}f_{3}f_{7}l_{4}\left(\tilde{c}-cf_{8}\right) & =0\,,\label{AdS2Hor_Bianchi3}\\
-bf_{5}^{2}f_{3}f_{4}l_{1}+\left(f_{5}^{4}l_{2}\right)'-f_{5}^{2}f_{3}f_{7}l_{4}\left(a-bf_{8}\right) & =0\,.\label{AdS2Hor_Bianchi4}
\end{align}
While $\eqref{AdS2Hor_Bianchi2}$ is satisfied automatically, the value of
$i_{2}$, the last unknown coefficient appearing in $G_{4}$ we are
left with, is determined by $\eqref{AdS2Hor_Bianchi3}$ and $\eqref{AdS2Hor_Bianchi4}$. Using $\eqref{AdS2Hor_KEqs}$ and defining
\[
\mathcal{I}\equiv\frac{i_{2}}{f_{1}^{2}}\,,
\]
we arrive at 
\begin{eqnarray}
3bK\mathcal{I}+F_{3}\mathcal{I}' & = & 0\label{AdS2Hor_IEq}\,,
\end{eqnarray}
which is solved by
\begin{equation}
\mathcal{I}=\frac{\mathcal{I}_{0}}{F_{3}^{3}}\,,\label{AdS2Hor_ISol}
\end{equation}
for some constant $\mathcal{I}_{0}\,$.

Thus, the only remaining equation is $\eqref{AdS2Hor_Bianchi1}$, which is a fourth-order nonlinear
ODE for $F_{5}$. Remarkably, this equation can be integrated twice into the second-order ODE:\footnote{The same technical simplification happens in the analysis of supersymmetric AdS$_3$ solution of type IIB supergravity in \cite{Benini:2015bwz}.}
\begin{equation}
\frac{1}{KF_{2}F_{3}^{2}}\left[-F_{5}'^{2}+2F_{5}\left(F_{5}''-2K\left(4qF_{2}F_{3}+\kappa F_{3}^{2}\right)\right)\right]-\Delta+{\cal P}_{3}=0\,,\label{AdS2Hor_F5Eq}
\end{equation}
where $\Delta(y)$ and ${\cal P}_{3}(y)$ satisfy the simple  differential equations:
\begin{align*}
\Delta'' & =24KF_{2}\frac{\mathcal{I}_{0}^{2}}{F_{3}^{4}}\,,\qquad\qquad
{\cal P}_{3}''  =32Kq\left(\kappa F_{3}+qF_{2}\right)\,.
\end{align*}
Using \eqref{F2F3app} and for $b\neq 0$ one may write  \eqref{AdS2Hor_F5Eq} as \eqref{AdS2Hor_F5EqMain}.

Although we have not found the most general solution to \eqref{AdS2Hor_F5Eq}, one can show that the most general polynomial solution is at most quartic, i.e.,  
\equ{\label{eq:F5polAPP}
F_{5}(y)=\sum_{n=0}^{4}\alpha_{n}\, y^{n}\,,
}
where $\alpha_n$ are real constants. Plugging this into \eqref{AdS2Hor_F5Eq} leads to a set of algebraic equations for the coefficients $\alpha_{n}$ and the other parameters specifying the solution:
\eqss{\label{constalphas}
6\alpha_{4}^{2}-3\alpha_{4}bK^{3}(b\kappa+4cq)+4b^{2}cK^{6}q(b\kappa+cq) & =0\,,\\ \\
+2K^{2}\left(2bK^{3}q\left(b^{2}\kappa S_{2}+5bc\kappa S_{3}+4bcqS_{2}+2c^{2}qS_{3}\right)-3\alpha_{4}(b\kappa S_{3}+2bqS_{2}+2cqS_{3})\right) \\
+3\alpha_{3}\left(3\alpha_{4}-bK^{3}(b\kappa+4cq)\right)& =0\,,\\ \\
\frac{12}{K}\alpha_{2}\left(3\alpha_{4}-bK^{3}(b\kappa+4cq)\right)+9\alpha_{3}^{2}-24\alpha_{3}K(b\kappa S_{3}+2bqS_{2}+2cqS_{3})\\
+4S_{3}\left(4c^{2}K^{3}q^{2}S_{3}-3\alpha_{4}(\kappa S_{3}+4qS_{2})\right)+b^{2}K^{3}(3cp_{1}+16qS_{2}(5\kappa S_{3}+3qS_{2}))\\
+16bcK^{3}qS_{3}(7\kappa S_{3}+8qS_{2}) & =0\,,\\ \\
4\alpha_{1}\left(4\alpha_{4}-bK^{3}(b\kappa+4cq)\right)+4\alpha_{2}\left(\alpha_{3}-2K^{2}(b\kappa S_{3}+2bqS_{2}+2cqS_{3})\right)\\
-4\alpha_{3}\kappa KS_{3}^{2}-16\alpha_{3}KqS_{3}S_{2}+b^{2}cK^{4}p_{0}+b^{2}K^{3}p_{1}S_{2}+2bcK^{3}p_{1}S_{3}\\
+\frac{112}{3}b\kappa K^{3}qS_{3}^{2}S_{2}+32bK^{3}q^{2}S_{3}S_{2}^{2}+16c\kappa K^{3}qS_{3}^{3}+\frac{64}{3}cK^{3}q^{2}S_{3}^{2}S_{2} & =0\,,\\ \\
-\frac{4}{K}\alpha_{0}b^{2}\left(bK^{3}(b\kappa+4cq)-6\alpha_{4}\right)-\frac{2}{K}\alpha_{1}b^{2}\left(4K^{2}(b\kappa S_{3}+2bqS_{2}+2cqS_{3})-3\alpha_{3}\right)\\
+b^{2}S_{3}(KS_{3}(cp_{1}+16qS_{2}(\kappa S_{3}+qS_{2}))-4\alpha_{2}(\kappa S_{3}+4qS_{2}))\\
+b^{4}K^{2}p_{0}S_{2}+2b^{3}KS_{3}(cKp_{0}+p_{1}S_{2})-12c^{2}K\text{\ensuremath{\mathcal{I}_{0}^{2}}} & =0\,,\\ \\
4\alpha_{0}b^{3}(\alpha_{2}-KS_{3}(\kappa S_{3}+4qS_{2}))+S_{2}\left(b^{3}Kp_{0}S_{3}^{2}-4bS_{2}\mathcal{I}_{0}^{2}-8cS_{3}\mathcal{I}_{0}^{2}\right)-\alpha_{1}^{2}b^{3} & =0\,,\\ \\
K\left(-4\alpha_{1}b^{3}S_{3}(\kappa S_{3}+4qS_{2})+2b^{4}Kp_{0}S_{3}S_{2}+b^{3}S_{3}^{2}(cKp_{0}+p_{1}S_{2})-16bcS_{2}\mathcal{I}_{0}^{2}-8c^{2}S_{3}\mathcal{I}_{0}^{2}\right)\\
-4\alpha_{0}b^{3}\left(2K^{2}(b\kappa S_{3}+2bqS_{2}+2cqS_{3})-3\alpha_{3}\right) & =0\,.
}
All the solutions discussed in this paper correspond to different solutions to this set of algebraic constraints. 

We note that assuming $b\neq 0$ the sixth equation above becomes
\[
\frac{cK\mathcal{I}_{0}^{2}(bS_{2}-cS_{3})}{S_{3}^{3}b}=0\,
\]
after solving the others. When instead $b=0$, one of the equations is not independent of the others. In both cases, the system leaves (at least) one free parameter.

%%%%%%%%%%%%%%%%%
\subsection{Singularity analysis}
\label{Appendix11D-Singularities}
%%%%%%%%%%%%%%%%%

The metric $\eqref{AdS2Hor_AltAnsatzMain}$ could in principle have conical
singularities when $y=\tilde{y}$ such that $F_{5}(\tilde{y})=0$.
In this section, we analyze what happens in these points and show that
they are regular. Let us consider the following linear combination
of the angles $\beta$ and $\psi$:
\[
\left(\begin{array}{c}
\beta\\
\psi
\end{array}\right)=W\left(\begin{array}{c}
\beta'\\
\psi'
\end{array}\right),\quad W=\left(\begin{array}{cc}
w_{1} & w_{2}\\
w_{3} & w_{4}
\end{array}\right),\quad W\mbox{ invertible}\,.
\]
Under this transformation the metric retains its structure, i.e.,
upon some redefinitions of the functions $F_{n}$, we can rewrite
it in the same form, with new parameters given by
\begin{equation}
\left(\begin{array}{c}
c\\
\tilde{c}
\end{array}\right)=W\left(\begin{array}{c}
c'\\
\tilde{c}'
\end{array}\right),\quad\left(\begin{array}{c}
b\\
a
\end{array}\right)=W\left(\begin{array}{c}
b'\\
a'
\end{array}\right).\label{AdS2Hor_paramsWT}
\end{equation}
 In particular, the part that involves $y$ and $\beta$, near $\tilde{y}$,
becomes
\[
ds_{y\beta}^{2}=K\frac{\sqrt[3]{F_{1}F_{2}^{2}F_{3}^{4}}}{\left(y-\tilde{y}\right)\tilde{F}_{5}'}\left(dy^{2}+\frac{\left(y-\tilde{y}\right)^{2}\left(\tilde{F}_{5}'\right)^{2}\det^{2}W}{\left(w_{4}\tilde{F}_{2}\tilde{F}_{3}^{2}-\tilde{F}_{6}w_{2}\right)^{2}}D\beta^{2}\right).
\]
Assuming $y>\tilde y$ and performing the change of variables $r^{2}\equiv2\left(y-\tilde{y}\right)$
and choosing $w_{4}=4\omega_{2}$, $w_{2}=-4\omega_{1}$ and $w_{1},w_{3}$
so that $\det W=1$ and using $\eqref{AdS2Hor_Solution-F6}$ we
obtain
\[
ds_{y\beta}^{2}=K\frac{\sqrt[3]{F_{1}F_{2}^{2}F_{3}^{4}}}{\tilde{F}_{5}'}\left(dr^{2}+r^{2}D\beta^{2}\right)\,.
\]
We thus see that there are no conical singularities, provided 
$\beta\in[0,2\pi]$.\footnote{Again, this is very similar to the regularity analysis of the solutions in \cite{Benini:2015bwz}.} Moreover, from $\eqref{AdS2Hor_paramsWT}$, we
see that, using $\eqref{AdS2Hor_abConstraint}$, we can set $\omega_{2}$
to zero by an appropriate  $W$.
%%%%%%%%%%%%%%%%%%%%%%%%%%%%%%%%%%%%%

%%%%%%%%%%%%%%%%%%%%%%%%%%%%%%%%%%%%%
\end{appendices}
%%%%%%%%%%%%%%%%%%%%%%%%%%%%%%%%%%%%%

\bibliography{AdS4-BH}
\bibliographystyle{JHEP}

\end{document}